\documentclass[12pt]{article}
\usepackage{csquotes}

\usepackage{epsfig}
\usepackage{subcaption}
\usepackage{graphicx}

\usepackage{verbatim}
\usepackage{amsmath}
\usepackage{amssymb}
\usepackage{color}
\usepackage{bm}  
\usepackage{hyperref}
 \usepackage{stackrel}
\usepackage{cite}
 
\usepackage{fullpage}

\def\##1{{\bf #1}}
\def\=#1{\underline{\underline #1}}
\def\~#1{{\tilde{\bf #1}}}

\def\eps{\varepsilon}
\def\epso{\eps_{\scriptscriptstyle 0}}
\def\lambdao{\lambda_{\scriptscriptstyle 0}}
\def\muo{\mu_{\scriptscriptstyle 0}}
\def\ko{k_{\scriptscriptstyle 0}}
\def\etao{\eta_{\scriptscriptstyle 0}}

\def\kL{k_{\rm L}}
\def\kR{k_{\rm R}}

\def\koa{\ko a}
\def\kLa{\kL a}
\def\kRa{\kR a}

\def\smn{_{\rm smn}}
\def\mn{_{\rm mn}}

\def\epsr{\eps_{\rm r}}
\def\mur{\mu_{\rm r}}
\def\nr{n_{\rm r}}
\def\etar{\eta_{\rm r}}
\def\etas{\eta_{\rm s}}
\def\gammas{\gamma_{\rm s}}

\def\les{\left[}
\def\ris{\right]}
\def\lec{\left\{}
\def\ric{\right\}}

\def\.{\mbox{ \tiny{$^\bullet$} }}

\def\ux{\hat{\#x}}
\def\uy{\hat{\#y}}
\def\uz{\hat{\#z}}

\def\ur{\hat{\#r}}
\def\uphi{\hat{\mbox{\boldmath$\phi$}}}
\def\utheta{\hat{\mbox{\boldmath$\theta$}}}

\def\pinm{\pi_{\rm n}^{\rm m}}
\def\taunm{\tau_{\rm n}^{\rm m}}
\def\one{^{(1)}}
\def\three{^{(3)}}

\def\Dmn{D_{\rm mn}}
\def\an{a_{\rm n}}
\def\bn{b_{\rm n}}
\def\cn{c_{\rm n}}
\def\dn{d_{\rm n}}

\def\fnone{f_{\rm 1n}}
\def\fntwo{f_{\rm 2n}}
\def\fnthree{f_{\rm 3n}}
\def\fnfour{f_{\rm 4n}}
\def\fnfive{f_{\rm 5n}}
\def\fnsix{f_{\rm 6n}}
\def\fnseven{f_{\rm 7n}}
\def\fneight{f_{\rm 8n}}

\def\gnone{g_{\rm 1n}}
\def\gntwo{g_{\rm 2n}}
\def\gnthree{g_{\rm 3n}}
\def\gnfour{g_{\rm 4n}}

\def\eomn{_{\stackrel[{\rm o}]{\rm e}{}{_{\rm mn}}}}
\def\smn{_{\rm smn}}

\def\Vint{{\cal V}_{\rm int}}

\def\Vext{{\cal V}_{\rm ext}}

\def\Va{{\cal V}_{\rm a}}
\def\Sa{{\cal S}_{\rm a}}
\def\Vso{{\cal V}_{\rm so}}

\def\Einc{\#E_{\rm inc}}
\def\Hinc{\#H_{\rm inc}}
\def\Esca{\#E_{\rm sca}}
\def\Hsca{\#H_{\rm sca}}
\def\Fsca{\#F_{\rm sca}}
\def\Eint{\#E_{\rm int}}
\def\Hint{\#H_{\rm int}}

\def\Thetasca{F_{\theta_{\rm sca}}}
\def\Phisca{F_{\phi_{\rm sca}}}

\def\inc{_{\rm inc}}
\def\sca{_{\rm sca}}

\def\ext{_{\rm ext}}
\def\abs{_{\rm abs}}

\def\as{a_{\rm s}}
\def\ap{a_{\rm p}}

\def\un{\hat{\#n}}
\def\ur{\hat{\#r}}

\def\ra{{\#r}_{\rm a}}
\def\ura{\hat{\#r}_{\rm a}}
\def\thetaa{\theta_{\rm a}}
\def\phia{\phi_{\rm a}}

\def\rso{{\#r}_{\rm so}}

\def\psis{\underline{\psi}^{\rm s}}
\def\psia{\underline{\psi}^{\rm a}}
\def\Psis{\Psi^{\rm s}}
\def\Psia{\Psi^{\rm a}}

\def\Qb{ Q_{\rm b}}
\def\Qf{ Q_{\rm f}}
\def\QD{ Q_{\rm D}}

\def\Qabs{Q\abs}

\begin{document}

\begin{center}
\textbf{van~de~Hulst essay: Geometric-phase portrayal of electromagnetic scattering by a  three-dimensional object in free space } \\

\textit{Akhlesh Lakhtakia\footnote{Email: akhlesh@psu.edu}}\\

{
The Pennsylvania State University, Department of Engineering Science and Mechanics, Nanoengineered Metamaterials Group,  University Park, PA 16802, USA}\\

\end{center}

\begin{abstract} 
The concept of geometric  phase was applied to initiate the geometric-phase portrayal of electromagnetic  scattering by a
three-dimensional object in free space.  Whereas the incident electromagnetic field is that of an arbitrarily polarized plane wave,
the direction-dependent far-zone scattering amplitude can be used to define 
 direction-dependent Stokes parameters for the scattered field. Both symmetric and asymmetric Poincar\'e spinors were formulated to characterize the polarization states of incident plane wave and the far-zone scattering amplitude, and two different geometric phases were defined therefrom. Density plots of
both geometric phases were calculated for five different
homogeneous isotropic spheres with different linear constitutive properties and boundary conditions: dielectric-magnetic spheres (non-dissipative and dissipative), impedance spheres, perfect electrically conducting spheres, charged dielectric-magnetic spheres, 
dielectric-magnetic spheres with topologically insulating surface states, and isotropic chiral spheres. The incident plane waves were taken to be linearly and circularly polarized, for the sake of illustration.
Numerical results revealed that geometric-phase density plots possess significantly richer  features than their counterparts for the differential scattering efficiency. The geometric-phase portrayals exhibit enhanced sensitivity to changes in the size and  composition of the scatterer, the boundary conditions,   and the incident polarization state, suggesting promise for inverse-scattering problems.

\end{abstract}

\section*{Prologue}

\begin{small}
\flushright It was a very good year, \\
 For me to be  born in fifty-seven. \\
 It was a very good year for rockets in space. \\
  As FORTRAN began to calculate, \\
  Light was conceived to spew forth from a laser, \\
 And van de Hulst published his magnum opus.\\
 All that took place in nineteen fifty-seven.\\
\end{small}
\vspace{1cm}

1957 was indeed a very good year. As the foregoing ditty (modeled on a song made famous in 1965 by Frank Sinatra) recalls, that year
two spacecraft were successfully launched \cite{Williamson}, the first commercial FORTRAN compiler was released \cite{Backus}, the laser was conceived \cite{Laser}, and
Hendrik Christoffel van~de~Hulst published his seminal book on light scattering by  particles \cite{vdH}.

But it was late in 1980, when both that book and I were 23 years old, that we met for the first time, while I was investigating
the plane-wave scattering matrix theory \cite{Kerns} for my doctoral research in bioelectromagnetics. Dr. van~de~Hulst's
magnum opus \cite{vdH} is dedicated to the scattering of plane waves by mostly three-dimensional (3D) objects, especially spheres, in line with his
 doctoral research  on light scattering by spheres in the astronomical context of interstellar dust \cite{vdH-PhD}. That research led
 to a landmark paper published in 1949, wherein he intuitively derived a relationship between the extinction cross section   and a
 specific component of the forward-scattered light by a 3D object illuminated by a plane wave \cite{vdH1949}.

Dr. van~de~Hulst's intuition was on the mark. Just six years later, the extinction cross section
was related exactly to the co-polarized scattering amplitude in the
forward direction by Jones \cite{JonesDS} and Saxon \cite{Saxon} independently.
Just three years later, de~Hoop \cite{deHoop} provided a derivation simpler
than that of Jones \cite{JonesDS} and equivalent to that of Saxon \cite{Saxon}. Variously called the cross-section theorem \cite{deHoop},
the extinction theorem \cite[p.~39]{vdH}, the optical theorem \cite{Jackson,BH83},
and the forward-scattering theorem \cite{Karam}, this relationship holds
true for co-moving observers but not for other inertial observers \cite{Garner}.

After using Ref.~\citenum{vdH} for over a dozen years, I had the great pleasure of meeting its author  in State College, PA,
at Prof. Craig Bohren's home in 1992. And it was an unexpected pleasure to be named a recipient of the 2025 van~de~Hulst Light-Scattering Award, which has led me to write this essay. My sense of symmetry compelled me to focus it on plane-wave scattering by 3D objects, with numerical results specialized to spheres.

A disclaimer: True plane waves exist only in books. A true plane wave: extends infinitely in space, containing infinite energy;
varies sinusoidally in time for all time; and has a fixed polarization state everywhere and for all time. Nevertheless, many observable phenomenons can be mathematically analyzed and explained quite easily using plane waves, the simplest of those phenomenons being the reflection and refraction of light by specularly smooth and planar bi-medium interfaces.

 \section{Introduction}\label{sec:intro}

The modest aim of this essay is to apply the concept of geometric phase, devised to compare
two plane waves, to the  far-zone scattered field arising from the plane-wave illumination of a 3D object in free space and thereby
initiate the geometric-phase portrayal of electromagnetic scattering.

Any uniform plane wave propagating in free space can be represented as a location on the surface of the Poincar\'e sphere
$s_1^2+s_2^2+s_3^2=s_0^2$, where $s_0$, $s_{1}$, $s_2$, and $s_3$ are the four Stokes parameters
of that plane wave \cite{BH83,Jackson}.
Two plane waves are dissimilar if their locations on the Poincar\'e sphere do not coincide, as was explicated by Pancharatnam in 1956
\cite{Pancha}, the dissimilarity being quantitated as the geometric phase.
Its profound roles in classical and quantum physics were recognized within the next three decades  \cite{SWbook}. Not only does
it continue to   fascinate researchers \cite{CLBNGK,Citro}, but it also finds applications  in the design
of planar devices such as achromatic phase shifters, spatial light modulators, frequency shifters, and planar lenses for wavefront engineering \cite{Brasselet2017,Kobashi2016,metasurface,Jisha,Faraz}.

No plane wave can be found in the scattered field when a 3D object is illuminated by an electromagnetic field emitted by a source.
Instead, in the region outside a sphere circumscribing that 3D object, the scattered field is a superposition of spherical waves \cite{FLbook,Kristensson,Wood,Lakh-EBCM}.

If the source is sufficiently far away from the 3D object such that the wavefront curvature becomes negligible, which requires that the distance between the source and the object to be much larger than the object's linear dimensions and the free-space wavelength $\lambdao$, then
the incident field can be reasonably approximated by a plane wave. By the same token, far away from the object in any direction, the scattered field is transversely polarized with respect to the scattering direction and can be considered as a plane wave whose electric and magnetic field phasors decay inversely with distance from the object. These two \textit{approximations} together underlie the plane-wave scattering matrix useful for radar \cite{VB,Graves,Huynen} and
also supply the rationale to apply the concept of geometric phase to 
electromagnetic scattering by a 3D object.

In the following sections, the  geometric-phase
portrayal of electromagnetic scattering is initiated and illustrated. Section~\ref{sec:bvp} describes the scattering boundary-value problem. Section~\ref{sec:sef} is devoted to the scattered electromagnetic field, first representing  the scattered field phasors in terms of vector spherical wavefunctions that are regular at infinity,
then determining a general expression for the far-zone scattering amplitude, and finally setting up direction-dependent Stokes parameters and two
Poincar\'e spinors of the far-zone scattered field. Section~\ref{sec:ief} first represents  the incident field phasors in terms of vector spherical wavefunctions that are regular at the origin,
then specializes that representation for an incident plane wave of arbitrary polarization state, and finally sets up  two
Poincar\'e spinors of the incident plane wave. Five standard measures of plane-wave scattering, such as the differential scattering efficiency and the backscattering efficiency, are reproduced in Sec.~\ref{sec:mpws1}. The symmetric and asymmetric geometric phases, both direction dependent,
of the far-zone scattered field are defined in Sec.~\ref{sec:gp}. Section~\ref{sec:results} provides 
illustrative examples of the geometric-phase portrayal of plane-wave scattering by an isotropic homogeneous sphere
for five different sets of linear constitutive properties and boundary conditions prevailing
 on the surface of the sphere. The essay concludes with a remark in Sec.~\ref{sec:cr}.

A note on notation: An  {$\exp(-i\omega{t})$} time dependence is implicit, with $i=\sqrt{-1}$, $\omega$ as the angular frequency, and $t$ as time.  The free-space wavenumber $\ko =2\pi/\lambdao= \omega \sqrt{\epso \muo}$ and  the free-space intrinsic impedance $\etao = \sqrt{\muo/\epso}$,
where $\epso$ and $\muo$ are the
permittivity and permeability of free space, respectively.
Vectors are displayed in bold typeface. The unit vectors
in the Cartesian coordinate system $(x,y,z)$ are denoted by $\ux$, $\uy$, and $\uz$.
The unit vectors
in the spherical coordinate system $(r,\theta,\phi)$ are denoted by $\ur$, $\utheta$, and $\uphi$. 
Column vectors are underlined and enclosed by square brackets. Dyadics are double underlined,
with $\=I=\ux\ux+\uy\uy+\uz\uz=\ur\ur+\utheta\utheta+\uphi\uphi$ representing the identity dyadic.
The superscript $^\ast$ indicates the complex conjugate and
the superscript $^\dag$ indicates the conjugate transpose of a column vector.

\section{{Boundary-Value Problem}}\label{sec:bvp}

\begin{figure}[htb]
      \centering
     \includegraphics[scale=0.4]{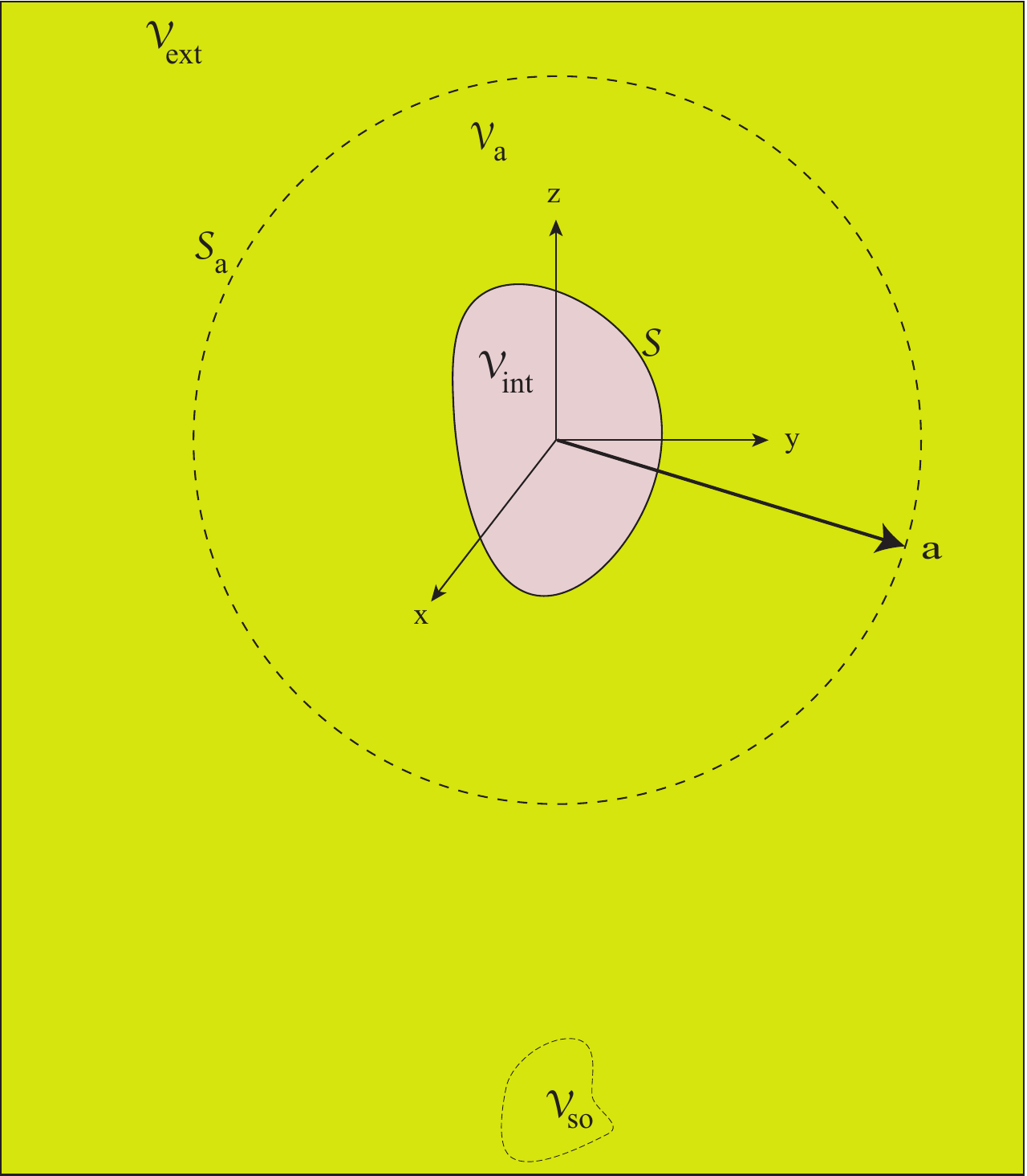}
      \vspace{0.5cm}
      \caption{Schematic illustrating the boundary-value problem.}
      \label{Fig1}
  \end{figure}

Let all space $\cal V$ be divided into two mutually disjoint  regions $\Vint$ and $\Vext$, as shown in Fig.~\ref{Fig1}.
The region $\Vint$ is bounded in all directions by the closed surface $\cal S$ and filled with a linear medium different from free space. Extending
to infinity in all directions, the
region $\Vext$ is vacuous. Furthermore, $\Vint  \subseteq\Va$, where $\Va$ is the sphere $r\leq{a}$, and the origin
of the  coordinate system is assumed to lie in the interior of $\Vint$.

The incident electromagnetic field is not necessarily a plane wave, but its sources 
are confined to a finite region $\Vso\subset \Vext$ that lies outside $\Va$
\cite{Wood,LI1983} and are assumed to be unaffected by the scattered electromagnetic field. 

The determination of the scattered field in terms of the incident field requires the formulation and solution of a boundary-value problem.
A host of analytical
and numerical techniques \cite{BSU,Lakh-MoM,Yurkin,Chew,SIE,Sertel,Lakh-EBCM,Vavilin,Elsherbeni,FEM-BEM,MAS} exist for that purpose. This essay is not concerned with those techniques, but with certain properties of the scattered field in relation to that of a plane wave illuminating $\Vint$. Therefore, it is
assumed that the scattered electric field phasor  $\Esca(\#r)$ and the scattered magnetic field phasor $\Hsca(\#r)$ are known for $\#r\in\Vext$.

\section{{Scattered Electromagnetic Field}}\label{sec:sef}

The direct sources of the
scattered field lie inside   $\Vint$. With  ${\Sa}$ denoting the exterior surface  of $\Va$,
 application of  the Huygens principle in the region ${\cal V}-\Va$ therefore delivers \cite[Sec.~9.8]{Chen}
\begin{subequations}
 \begin{eqnarray}
\nonumber
\Esca(\#r)&=&\iint_{\Sa} \lec
\les\nabla\times\=G(\#r,\ra)\ris\.\les \un(\ra)\times\Esca(\ra)\ris
\right.
\\[8pt]
&&+\left.
i\omega\muo \,\=G(\#r,\ra)\.\les \un(\ra)\times\Hsca(\ra)\ris
\ric d^2\ra\,,\quad {r>a},
\label{def-Esca}
\end{eqnarray}
and
 \begin{eqnarray}
\nonumber
\Hsca(\#r)&=&\iint_{\Sa} \lec
\les\nabla\times\=G(\#r,\ra)\ris\.\les \un(\ra)\times\Hsca(\ra)\ris
\right.
\\[8pt]
&&-\left.
i\omega\epso \,\=G(\#r,\ra)\.\les \un(\ra)\times\Esca(\ra)\ris
\ric d^2\ra\,,\quad {r>a},
\label{def-Hsca}
\end{eqnarray}
\end{subequations}
where $\un(\ra)$ is the unit   normal to $\Sa$ at $\ra\in\Sa$ and points into
${\cal V}-\Va$. The  dyadic Green function for free space 
\begin{equation}
\=G(\#r,\ra)= \left(\=I + \frac{\nabla\nabla}{\ko^2}\right)\frac{\exp\left(i\ko\vert\#r-\ra\vert\right)}{4\pi\vert\#r-\ra\vert}
\end{equation}
is used because the region ${\cal V}-\Va$ is vacuous.
The radiation condition at infinity has been used in the derivation of Eqs.~(\ref{def-Esca}) and (\ref{def-Hsca}).

\subsection{{Series Representation of Scattered Field Phasors}}

The bilinear form \cite{FLbook}
\begin{eqnarray}
\=G(\#r,\ra)&=& \frac{i\ko}{\pi}
\sum_{s\in\left\{e,o\right\}}\sum^{\infty}_{n=1}\sum^n_{m=0}\Dmn\les
{{\#M}\smn\three}(\ko\#r) {{\#M}\smn\one}(\ko\ra)
+ {{\#N}\smn\three}(\ko\#r) {{\#N}\smn\one}(\ko\ra)\ris
\label{G-bilinear-1}
\end{eqnarray}
can be used in Eqs.~(\ref{def-Esca}) and (\ref{def-Hsca})
because $r>a$ and $\vert\ra\vert=a$. Here,
the normalization factor
\begin{equation}
\displaystyle{
\Dmn=(2-\delta_{\rm m0}){(2n+1)(n-m)!\over 4n(n+1)(n+m)!}
}
\end{equation}
employs the Kronecker delta $\delta_{\rm mm^\prime}$ and
 the vector spherical wavefunctions ${{\#M}\smn^{(j)}}(\ko\#r)$ and ${{\#N}\smn^{(j)}}(\ko\#r)$    {\cite{Stratton1941,MF1953}} are defined in   Appendix~1.
 
 Substitution of Eq.~(\ref{G-bilinear-1}) in Eq.~(\ref{def-Esca}) leads to
 \begin{eqnarray}
 \Esca(\#r)&=& \sum_{s\in\left\{e,o\right\}}\sum^{\infty}_{n=1}\sum^n_{m=0}\Dmn\les
 A\smn\three\,{{\#M}\smn\three}(\ko\#r)
 \right.
 \left.
 +  B\smn\three\,{{\#N}\smn\three}(\ko\#r)\ris\,,\,\,r>a\,,
 \label{Esca}
 \end{eqnarray}
 where the scattered field coefficients
 \begin{eqnarray}
\nonumber
 A\smn\three&=&\frac{i\left(\ko a\right)^2}{\pi}\int_{\phia=0}^{2\pi}\int_{\thetaa=0}^{\pi}
  \lec
\#N\smn\one(\ko a\ura)\.\les \ura\times\Esca(a\ura)\ris
\right.
\\[8pt]
&&+\left.
i\etao \,\#M\smn\one(\ko a\ura)\.\les \ura\times\Hsca(a\ura)\ris
\ric \sin\thetaa\,d\thetaa\,d\phia\,
\label{def-Asmn3}
\end{eqnarray}
and
 \begin{eqnarray}
\nonumber
B\smn\three&=&\frac{i\left(\ko a\right)^2}{\pi}\int_{\phia=0}^{2\pi}\int_{\thetaa=0}^{\pi}\lec
\#M\smn\one(\ko a\ura)\.\les \ura\times\Esca(a\ura)\ris
\right.
\\[8pt]
&&+\left.
i\etao\, \#N\smn\one(\ko a\ura)\.\les \ura\times\Hsca(a\ura)\ris
\ric \sin\thetaa\,d\thetaa\,d\phia\,
\label{def-Bsmn3}
\end{eqnarray}
employ the unit vector $\ura=\left(\ux\cos\phia+\uy\sin\phia\right)\sin\thetaa+\uz\cos\thetaa$.
Likewise, substitution of Eq.~(\ref{G-bilinear-1}) in Eq.~(\ref{def-Hsca})
delivers
 \begin{eqnarray}
 \Hsca(\#r)&=&-i\etao^{-1} \sum_{s\in\left\{e,o\right\}}\sum^{\infty}_{n=1}\sum^n_{m=0}\Dmn\les
 A\smn\three\,{{\#N}\smn\three}(\ko\#r)
 \right.
 \left.
 +  B\smn\three\,{{\#M}\smn\three}(\ko\#r)\ris\,,\,\,r>a\,.
 \label{Hsca}
 \end{eqnarray}
 
 \subsection{{Far-Zone Scattering Amplitude}}\label{sec:ffsa}
 In the far zone (i.e., as $\ko r\to \infty$), the scattered electric field  phasor may be approximated as \cite{Saxon,Kristensson}
\begin{subequations}
\begin{equation}
\Esca(r\hat{\#r})\approx\#F\sca(\hat{\#r}) \frac{\exp(i{\ko}r)}{r}\,
\label{def-Esca-far}
\end{equation}
and the scattered magnetic field phasor as
\begin{equation}
\Hsca(r\hat{\#r})\approx\etao^{-1}\hat{\#r}\times\#F\sca(\hat{\#r}) \frac{\exp(i{\ko}r)}{r}\,,
\label{def-Hsca-far}
\end{equation}
\end{subequations}
where  $\hat{\#r}=\left(\ux\cos\phi+\uy\sin\phi\right)\sin\theta+\uz\cos\theta$. The
  far-zone scattering amplitude
\begin{eqnarray}
\#F\sca(\hat{\#r}) = \frac{1}{\ko} \les\Thetasca(\hat{\#r}) \utheta
 +\Phisca(\hat{\#r}) \uphi\ris\,
\label{def-Fsca}
\end{eqnarray}
does not have a component parallel to $\ur$, with
\begin{subequations}
\begin{eqnarray}
&& \Thetasca(\hat{\#r}) 
=\sum_{s\in\left\{e,o\right\}}\sum^{\infty}_{n=1}\sum^n_{m=0}\lec
(-i)^n\,D\mn \les-iA\smn\three\,{f\smn}(\theta,\phi)
+B\smn\three\,{g\smn}(\theta,\phi)\ris \ric
\label{def-Thetasca}
\end{eqnarray}
and
\begin{eqnarray}
&&
 \Phisca(\hat{\#r}) 
=\sum_{s\in\left\{e,o\right\}}\sum^{\infty}_{n=1}\sum^n_{m=0}\lec
(-i)^n\,D\mn \les iA\smn\three\,{g\smn}(\theta,\phi)
+B\smn\three\,{f\smn}(\theta,\phi)\ris \ric\,
\label{def-Phisca}
\end{eqnarray}
\end{subequations}
involving the functions $f\smn(\theta,\phi)$ and $g\smn(\theta,\phi)$
 defined in   Appendix~1. 
 
 Only the terms with $m=1$ in Eqs.~(\ref{def-Thetasca})
 and (\ref{def-Phisca}) survive when $\theta\in\left\{0,\pi\right\}$, as shown in
 Appendix~2, so that
\begin{eqnarray}
&&\Fsca(\uz)=\frac{1}{4\ko}\sum_{n=1}^{\infty} 
\lec
 i^{-n} \frac{2n+1}{n(n+1)}
\les\left(-iA_{\rm o1n}\three+B_{\rm e1n}\three\right)\ux
+\left(iA_{\rm e1n}\three+B_{\rm o1n}\three\right)\uy\ris\ric
\label{def-Fsca-f}
 \end{eqnarray}
 and
\begin{eqnarray}
&&
\Fsca(-\uz)=\frac{1}{4\ko}\sum_{n=1}^{\infty}
\lec i^n \frac{2n+1}{n(n+1)}
\les-\left(iA_{\rm o1n}\three+B_{\rm e1n}\three\right)\ux
+\left(iA_{\rm e1n}\three-B_{\rm o1n}\three\right)\uy\ris\ric\,.
\label{def-Fsca-b}
 \end{eqnarray}
 Equations~(\ref{def-Fsca-f}) and (\ref{def-Fsca-b}) hold regardless of the spatial profile of the incident electromagnetic
 field.
 
 \subsection{Direction-dependent Stokes parameters of the far-zone scattered field}
According to Eqs.~(\ref{def-Esca-far}) and (\ref{def-Hsca-far}),   
$\Esca(r\hat{\#r})$ and $\Hsca(r\hat{\#r})$ are mutually transverse in the far zone. Furthermore,
according to Eq. (\ref{def-Fsca}), both $\Esca(r\hat{\#r})$ and $\Hsca(r\hat{\#r})$ are  transverse
to the scattering direction $\ur$. We can therefore think of the far-zone scattered field  as a plane wave
propagating in the scattering direction, albeit with field amplitudes that are inversely proportional to the 
propagation distance and depend on the scattering direction. This allows for the prescription of the four  Stokes parameters of the far-zone scattered 
field as follows \cite[Sec.~3.3]{BH83}:
\begin{subequations}
\begin{eqnarray}
&&s_{0\sca}(\ur)
=  \vert {\Thetasca(\ur)}\vert^2+ \vert {\Phisca(\ur)}\vert^2\,,
\\[5pt]
&&
s_{1\sca}(\ur) =  \vert {\Thetasca(\ur)}\vert^2- \vert {\Phisca(\ur)}\vert^2\,,
\\[5pt]
&&
s_{2\sca}(\ur) 
=-2\,{\rm Re}\les   {\Phisca(\ur) \,\Thetasca^\ast(\ur)}\ris\,,
\\[5pt]
&&
s_{3\sca}(\ur) 
=-2\,{\rm Im}\les   {\Phisca(\ur)\, \Thetasca^\ast(\ur)}\ris\,.
\end{eqnarray}
\end{subequations}
These Stokes parameters are direction dependent.

\subsection{Direction-dependent Poincar\'e spinors of the far-zone scattered field}
We collectively identify the four direction-dependent Stokes parameters on the Poincar\'e sphere by
 the longitude $\alpha\sca(\ur)\in[0,2\pi)$ and the latitude $\beta\sca(\ur)\in[-\pi/2,\pi/2]$ defined through the relations
\begin{equation}
\label{def-alphabeta}
\left.\begin{array}{l}
s_{1\sca}(\ur)=s_{0\sca}(\ur)\, \cos\beta\sca(\ur)\, \cos\alpha\sca(\ur)
\\[5pt]
s_{2\sca}(\ur)=s_{0\sca}(\ur)\, \cos\beta\sca(\ur)\, \sin\alpha\sca(\ur)
\\[5pt]
s_{3\sca}(\ur)=s_{0\sca}(\ur)\, \sin\beta\sca(\ur) 
\end{array}
\right\}\,.
\end{equation}

The  angles $\alpha\sca(\ur)$ and $\beta\sca(\ur)$ appear in a  direction-dependent Poincar\'e spinor that can have (at least) two different formulations. The symmetric Poincar\'e spinor  \cite{Gori}
\begin{equation}
\label{def-PS-sym}
\les{\psis\sca}(\ur)\ris=\les
\begin{array}{c}
\cos\les\frac{1}{4}\pi-\frac{1}{2}\beta\sca(\ur)\ris\exp\les -\frac{1}{2}i\alpha\sca(\ur)\ris
\\[5pt]
\sin\les\frac{1}{4}\pi-\frac{1}{2}\beta\sca(\ur)\ris\exp\les \frac{1}{2}i\alpha\sca(\ur)\ris
\end{array}
\ris\,
\end{equation}
is modeled after the Jones vector \cite{JonesRC}, whereas the asymmetric Poincar\'e spinor  \cite{Akash}
\begin{equation}
\label{def-PS-asym}
\les {\psia\sca}(\ur)\ris= \exp\les \frac{1}{2}i\alpha\sca(\ur)\ris\,\les{\psis\sca}(\ur)\ris
=\les
\begin{array}{c}
\cos\les\frac{1}{4}\pi-\frac{1}{2}\beta\sca(\ur)\ris
\\[5pt]
\sin\les\frac{1}{4}\pi-\frac{1}{2}\beta\sca(\ur)\ris\exp\les i\alpha\sca(\ur)\ris
\end{array}
\ris 
\end{equation}
is not.

\section{Incident Electromagnetic Field}\label{sec:ief}
The indirect sources of the scattered field lie inside $\Vso$. These sources are a source electric current density phasor $\#J_e(\#r)$ and a source magnetic current density phasor $\#J_m(\#r)$, which emit the  electric field phasor
\begin{subequations}
\begin{equation}
\#E\inc(\#r)=\int_{\Vso} \lec i\omega\muo\=G(\#r,\rso)\. \#J_e(\rso)-\les\nabla\times
\=G(\#r,\rso)\ris\.\#J_m(\rso)\ric\,d^3\rso\,,\quad \#r\notin\Vso\,,
\label{def-Einc-so}
\end{equation}
and the   magnetic field phasor
\begin{equation}
\#H\inc(\#r)=\int_{\Vso} \lec i\omega\epso\=G(\#r,\rso)\. \#J_m(\rso)+\les\nabla\times
\=G(\#r,\rso)\ris\.\#J_e(\rso)\ric\,d^3\rso\,,\quad \#r\notin\Vso\,,
\label{def-Hinc-so}
\end{equation}
\end{subequations}
incident on the 3D object. The process of scattering by the matter occupying $\Vint$ then generates the scattered field in $\Vext$.

Equations~(\ref{def-Einc-so}) and (\ref{def-Hinc-so}) suffice to formulate and solve the boundary-value problem using a host of numerical techniques \cite{Lakh-MoM, Yurkin,Chew,SIE,Sertel,Vavilin,Elsherbeni,FEM-BEM,MAS}. 

\subsection{{Series Representations of Incident Field Phasors}}
For some other techniques, especially the extended boundary condition method  \cite{LI1983,Lakh-EBCM}, the incident field needs to be represented inside $\Va$ in terms of vector spherical wavefunctions as 
\begin{subequations}
\begin{eqnarray}
&&
\Einc(\#r) = \sum_{s\in\left\{e,o\right\}}\sum^{\infty}_{n=1}\sum^n_{m=0}
\left\{D\mn\left[A\smn\one\,
{{\#M}\smn\one}(\ko\#r)
+B\smn\one\,
{{\#N}\smn\one}(\ko\#r)\right]\right\},\quad r<a\,,
\label{incE2}
\end{eqnarray}
and
\begin{eqnarray}
\nonumber
&& \Hinc(\#r) = -i\etao^{-1} \sum_{s\in\left\{e,o\right\}}\sum^{\infty}_{n=1}\sum^n_{m=0}
\left\{D\mn\left[A\smn\one\,
{{\#N}\smn\one}(\ko\#r)
\right.\right.\\
&&\qquad\qquad\qquad\left.\left.+
B\smn\one\,
{{\#M}\smn\one}(\ko\#r)\right]\right\},\quad r<a\,.
\label{incH2}
\end{eqnarray}
\end{subequations}

The incident field coefficients $A\smn\one$ and $B\smn\one$ have to be determined by applying the Huygens principle in $\Va$ \cite[Sec.~9.8]{Chen}. By definition, the incident field exists everywhere in ${\cal V}-\Vso$ when the region $\Vint$ is also vacuous. Therefore, the Huygens principle yields
\begin{subequations}
 \begin{eqnarray}
\nonumber
\Einc(\#r)&=&-\iint_{\Sa} \lec
\les\nabla\times\=G(\#r,\ra)\ris\.\les \un(\ra)\times\Einc(\ra)\ris
\right.
\\[8pt]
&&+\left.
i\omega\muo \,\=G(\#r,\ra)\.\les \un(\ra)\times\Hinc(\ra)\ris
\ric d^2\ra\,,\quad {r<a},
\label{def-Einc}
\end{eqnarray}
and
 \begin{eqnarray}
\nonumber
\Hinc(\#r)&=&-\iint_{\Sa} \lec
\les\nabla\times\=G(\#r,\ra)\ris\.\les \un(\ra)\times\Hinc(\ra)\ris
\right.
\\[8pt]
&&-\left.
i\omega\epso \,\=G(\#r,\ra)\.\les \un(\ra)\times\Einc(\ra)\ris
\ric d^2\ra\,,\quad {r<a}.
\label{def-Hinc}
\end{eqnarray}
\end{subequations}
The bilinear form \cite{FLbook}
\begin{eqnarray}
\=G(\#r,\ra)&=& \frac{i\ko}{\pi}
\sum_{s\in\left\{e,o\right\}}\sum^{\infty}_{n=1}\sum^n_{m=0}\Dmn\les
{{\#M}\smn\one}(\ko\#r) {{\#M}\smn\three}(\ko\ra)
+ {{\#N}\smn\one}(\ko\#r) {{\#N}\smn\three}(\ko\ra)\ris
\label{G-bilinear-2}
\end{eqnarray}
can be used on the right sides of Eqs.~(\ref{def-Einc}) and (\ref{def-Hinc})
because $r<\vert\ra\vert=a$. Substitution of Eq.~(\ref{G-bilinear-2}) on the right sides
of Eqs.~(\ref{def-Einc}) and (\ref{def-Hinc}) and comparison with the right sides
of Eqs.~(\ref{incE2}) and (\ref{incH2}) deliver the incident field coefficients
\begin{eqnarray}
\nonumber
 A\smn\three&=&-\frac{i\left(\ko a\right)^2}{\pi}\int_{\phia=0}^{2\pi}\int_{\thetaa=0}^{\pi}
  \lec
\#N\smn\three(\ko a\ura)\.\les \ura\times\Einc(a\ura)\ris
\right.
\\[8pt]
&&+\left.
i\etao\, \#M\smn\three(\ko a\ura)\.\les \ura\times\Hinc(a\ura)\ris
\ric \sin\thetaa\,d\thetaa\,d\phia\,
\label{def-Asmn1}
\end{eqnarray}
and
 \begin{eqnarray}
\nonumber
B\smn\three&=&-\frac{i\left(\ko a\right)^2}{\pi}\int_{\phia=0}^{2\pi}\int_{\thetaa=0}^{\pi}\lec
\#M\smn\three(\ko a\ura)\.\les \ura\times\Einc(a\ura)\ris
\right.
\\[8pt]
&&+\left.
i\etao\, \#N\smn\three(\ko a\ura)\.\les \ura\times\Hinc(a\ura)\ris
\ric \sin\thetaa\,d\thetaa\,d\phia\,.
\label{def-Bsmn1}
\end{eqnarray}
Equations~(\ref{def-Einc-so}) and (\ref{def-Hinc-so}) have to be used on the
right sides of Eqs.~(\ref{def-Asmn1}) and (\ref{def-Bsmn1}).

\subsection{Incident Plane Wave}

As discussed in Sec.~\ref{sec:intro}, the concept of the geometric phase was devised to compare the locations of two plane waves on the Poincar\'e sphere. In the present context,
one of those two plane waves can be replaced by $\Fsca(\ur)$. The other plane wave
should be independent of $\ur$ in order to serve as a reference for all $\ur$; that role
can be fulfilled by the field emitted by the sources in $\Vso$, provided that $\Vso$ is sufficiently far away from $\Vint$. Therefore, the illumination of a 3D object   by a plane wave is considered in the remainder of this essay.

Without any loss of generality, the $z$ axis can be chosen such that the incident plane wave propagates in the
direction of increasing $z$. The electric field phasor of this plane wave can be written as
\begin{subequations}
\begin{equation}
\label{Einc-plw}
\Einc(\#r)=\left(\as \ux+  \ap\uy\right)\exp(i\ko z)\,,
\end{equation}
and the magnetic field phasor as
\begin{equation}
\label{Hinc-plw}
\Hinc(\#r)=\etao^{-1}\,\left(\as \uy-  \ap\ux\right)\exp(i\ko z)\,,
\end{equation}
\end{subequations}
where the coefficients $\as\in\mathbb{C}$ and $\ap\in\mathbb{C}$. The incident plane wave
is: right elliptically polarized if ${\rm Im}(\as\ap^\ast)>0$, 
left elliptically polarized if  ${\rm Im}(\as\ap^\ast)<0$, and
linearly polarized  if ${\rm Im}(\as\ap^\ast)=0$.

The incident field coefficients appearing in Eqs.~(\ref{incE2}) and (\ref{incH2}) can be obtained as
\cite[Eq. (13.3.70)]{MF1953}
\begin{subequations}
\begin{equation}
\label{A1-plw}
A\smn\one=-2n(n+1)i^n\left(\ap\,\delta_{se}-\as\delta_{so}\right)\delta_{m1}
\end{equation}
and
\begin{equation}
\label{B1-plw}
B\smn\one=-2n(n+1)i^{n+1}\left(\as\,\delta_{se}+\ap\delta_{so}\right)\delta_{m1}\,.
\end{equation}
\end{subequations}

\subsection{Poincar\'e Spinors of Incident Plane Wave}
The four Stokes parameters of the incident plane wave are as follows:
\begin{subequations}
\begin{eqnarray}
&&s_{0\inc}=  \vert\as\vert^2+\vert\ap\vert^2\,,
\\[5pt]
&&s_{1\inc}=  \vert\ap\vert^2-\vert\as\vert^2\,,
\\[5pt]
&&s_{2\inc}=2{\rm Re}\left(\as\ap^\ast\right)\,,
\\[5pt]
&&s_{3\inc}=2{\rm Im}\left(\as\ap^\ast\right)\,.
\end{eqnarray}
\label{Stokes-inc}
\end{subequations}
These
Stokes parameters are collectively identified by
 the longitude $\alpha\inc \in[0,2\pi)$ and the latitude $\beta\inc \in[-\pi/2,\pi/2]$ defined through the relations
\begin{equation}
\label{def-alphabeta-inc}
\left.\begin{array}{l}
s_{1\inc}=s_{0\inc} \cos\beta\inc \, \cos\alpha\inc 
\\[5pt]
s_{2\inc}=s_{0\inc}\cos\beta\inc\, \sin\alpha\inc 
\\[5pt]
s_{3\inc} =s_{0\inc} \sin\beta\inc 
\end{array}
\right\}\,.
\end{equation}
Hence,
\begin{equation}
\label{def-PS-sym-inc}
\les{\psis\inc}\ris=\les
\begin{array}{c}
\cos\left(\frac{1}{4}\pi-\frac{1}{2}\beta\inc\right)\exp\left( -\frac{1}{2}i\alpha\inc\right)
\\[5pt]
\sin\left(\frac{1}{4}\pi-\frac{1}{2}\beta\inc\right)\exp\left( \frac{1}{2}i\alpha\inc\right)
\end{array}
\ris\,
\end{equation}
is   the symmetric  Poincar\'e spinor  and
\begin{equation}
\label{def-PS-asym-inc}
\les{\psia\inc} \ris= \exp\left(\frac{1}{2}i\alpha\inc\right)\les{\psis\inc}\ris=
\les
\begin{array}{c}
\cos\left(\frac{1}{4}\pi-\frac{1}{2}\beta\inc\right)
\\[5pt]
\sin\left(\frac{1}{4}\pi-\frac{1}{2}\beta\inc\right) \exp\left(i\alpha\inc \right)
\end{array}
\ris\,
\end{equation}
is the asymmetric  Poincar\'e spinor  
of the incident plane wave.

\subsubsection{Linearly polarized incident plane wave}
With  $\ap=0$, we get
$\alpha\inc=\pi$ and $\beta\inc=0$  from Eqs.~(\ref{Stokes-inc}) and (\ref{def-alphabeta-inc}) so that
\begin{subequations}
\begin{equation}
\les {\psis\inc }\ris=\frac{1}{\sqrt{2}}\les \begin{array}{c} -i\\ i\end{array}\ris\,
\end{equation}
and
\begin{equation}
\les {\psia\inc }\ris=\frac{1}{\sqrt{2}}\les \begin{array}{c} 1\\-1\end{array}\ris\,
\end{equation}
\end{subequations}
for an $s$-polarized incident plane wave. Likewise,
on setting $\as=0$,   we get
$\alpha\inc=0$, $\beta\inc=0$, and
\begin{equation}
\les {\psis\inc }\ris=
\les {\psia\inc} \ris=\frac{1}{\sqrt{2}}\les \begin{array}{c} 1\\1\end{array}\ris\,
\end{equation}
for a $p$-polarized incident plane wave.

\subsubsection{Circularly polarized incident plane wave}
With $\as=i\ap$, Eqs.~(\ref{Stokes-inc}) and (\ref{def-alphabeta-inc}) 
deliver
$\alpha\inc=0$ and $\beta\inc=\pi/2$, so that
\begin{equation}
\les {\psis\inc }\ris=
\les{\psia\inc} \ris= \les \begin{array}{c} 1\\0\end{array}\ris\,
\end{equation}
is the Poincar\'e spinor of a right-circularly polarized (RCP) incident plane wave.
On setting
$\as=-i\ap$ in the same equations, it follows that $\alpha\inc=0$, $\beta\inc=-\pi/2$, and
\begin{equation}
\les {\psis\inc }\ris=
\les{\psia\inc} \ris= \les \begin{array}{c} 0\\1\end{array}\ris\,
\end{equation}
for a left-circularly polarized (LCP) incident plane wave.

\section{Standard Measures of Plane-Wave Scattering}\label{sec:mpws1}
Scattering being different in different directions, the chief standard measure of plane-wave scattering by a 3D object is
the differential scattering efficiency \cite{Saxon}
\begin{equation}\label{sigmad-def}
\QD(\ur)\triangleq\frac{4}{a^2} 
\frac{\#F\sca(\ur)\.\#F\sca^\ast(\ur)}{\vert\as\vert^2+\vert\ap\vert^2} \,
\end{equation}
along any radial direction specified by the unit vector $\ur\equiv(\theta,\phi)$. 
It is proportional to the bistatic radar cross-section \cite{RCS}.

The total scattering efficiency  is  calculated as  the double integral
\begin{equation} 
\label{tse-def0}
Q\sca=
\frac{1}{4\pi}\,\int_{\phi=0}^{2\pi}\int_{\theta=0}^{\pi}\,\QD(\theta,\phi)\sin\theta\,d\theta\,d\phi\,,
\end{equation}
which yields
 \begin{equation} 
\label{tse}
Q\sca=
\displaystyle{
\frac{1}{\ko^2a^2} \frac
{\displaystyle{\sum_{s\in\left\{e,o\right\}}\sum^{\infty}_{n=1}\sum^n_{m=0}}
\les D\mn\left(\vert A\smn\three\vert^2+\vert B\smn\three\vert^2\right)\ris}
{\vert\as\vert^2+\vert\ap\vert^2}
}\,
\end{equation}
upon the use of Eqs.~(\ref{def-Fsca}), (\ref{def-Thetasca}), and (\ref{def-Phisca}).
The extinction efficiency is defined as  {\cite{deHoop,Saxon}}
\begin{equation}
\label{Qext-def0}
Q\ext=\frac{4}{\ko{a^2}} 
\frac{{\rm Im} \lec{\#F\sca(\uz)\.\left(\as^\ast\ux+\ap^\ast\uy\right)}\ric}{\vert\as\vert^2+\vert\ap\vert^2} \,.
\end{equation}
Substitution of  Eq.~(\ref{def-Fsca-f}) in this definition delivers
 \begin{equation}
 Q\ext=\frac{1}{\ko^2a^2} {\rm Im}\left(
 \sum_{n=1}^{\infty}\lec
 i^{-n} \frac{2n+1}{n(n+1)}
 \les\frac{\left(-iA_{\rm o1n}\three+B_{\rm e1n}\three\right)\as^\ast+
\left(iA_{\rm e1n}\three+B_{\rm o1n}\three\right)\ap^\ast}
{\vert\as\vert^2+\vert\ap\vert^2}
\ris
 \ric\right)\,.
 \end{equation}
 By virtue of the principle of conservation of energy,
the absorption efficiency $Q\abs=Q\ext-Q\sca$ must be non-negative.
 
The forward-scattering  efficiency  {\cite{BSU,Saxon}}
\begin{eqnarray}
\label{Qf-def0}
\Qf&=&\QD(\uz)
\\
\nonumber
&=&\frac{1}{4\ko^2a^2} \Bigg\{
\frac
{\left \vert  \displaystyle{\sum_{n=1}^{\infty}}\les
 i^{-n} \frac{2n+1}{n(n+1)}
\left(-iA_{\rm o1n}\three+B_{\rm e1n}\three\right)\ris\right\vert^2}
{\vert\as\vert^2+\vert\ap\vert^2}
\\[8pt]
&&
+
\frac{
\left \vert  \displaystyle{\sum_{n=1}^{\infty}}\les
 i^{-n} \frac{2n+1}{n(n+1)}
\left(iA_{\rm e1n}\three+B_{\rm o1n}\three\right)\ris\right\vert^2}
{\vert\as\vert^2+\vert\ap\vert^2}
\Bigg\}
\end{eqnarray}
is primarily of interest in certain effective-medium theories for particulate composite materials \cite{LOCM,WT,VMV}.
Finally, the backscattering efficiency  {\cite{BSU}}
\begin{eqnarray}
\label{Qb-def0}
\Qb&=&\QD(-\uz)
\\
\nonumber
&=&\frac{1}{4\ko^2a^2} \Bigg\{
\frac
{\left \vert  \displaystyle{\sum_{n=1}^{\infty}}\les
 i^{n} \frac{2n+1}{n(n+1)}
\left(iA_{\rm o1n}\three+B_{\rm e1n}\three\right)\ris\right\vert^2}
{\vert\as\vert^2+\vert\ap\vert^2}
\\[8pt]
&&
+
\frac{
\left \vert  \displaystyle{\sum_{n=1}^{\infty}}\les
 i^{n} \frac{2n+1}{n(n+1)}
\left(iA_{\rm e1n}\three-B_{\rm o1n}\three\right)\ris\right\vert^2}
{\vert\as\vert^2+\vert\ap\vert^2}
\Bigg\}
 \end{eqnarray}
finds use through the monostatic radar cross-section \cite{RCS}.

\section{{Geometric Phases of Far-Zone Scattered   Field}}\label{sec:gp}
 
With respect to the incident plane wave,
the symmetric geometric phase of the far-zone scattered field in the direction $\ur$  is defined
as the angle
\begin{equation}
\label{def-GP-symm}
\Psis\sca(\ur)={\rm Arg}\lec {\les{\psis\inc}\ris}^\dag\cdot{\les{\psis\sca}(\ur)\ris}\ric\,,
\end{equation}
and
the asymmetric geometric phase of the far-zone scattered field in the direction $\ur$  
as the angle
\begin{equation}
\label{def-GP-asymm}
\Psia\sca(\ur)={\rm Arg}\lec {\les{\psia\inc}\ris}^\dag\cdot{\les{\psia\sca}(\ur)\ris}\ric\,.
\end{equation}
In view of Eqs.~(\ref{def-PS-asym}) and (\ref{def-PS-asym-inc}), the symmetric and asymmetric geometric phases are related as follows:
\begin{equation}
\Psia\sca(\ur)= \frac{\alpha\sca(\ur)-\alpha\inc}{2}+ \Psis\sca(\ur)\,.
\end{equation}

\textit{Aside}: In the context of planar optics, some researchers subtract the so-called dynamic phase \cite{Gutie2011} from   $\Psis\sca$ and call only the remainder as the geometric phase. The dynamic phase is the phase acquired by a plane wave as it propagates through a planar device from an entry plane to an exit plane, both planes being parallel to one another. This dynamic phase is generally considered equal to $\ko \bar{n}L$, where $\bar{n}$ is an effective refractive index and $L$ is the distance between the entry and the exit planes. Whereas $\bar{n}$ is easy to define if the material between the two planes is isotropic,  achiral, and homogeneous, that quantity is not amenable to unimpeachable definition
if otherwise (e.g., if the material is   anisotropic dielectric or isotropic chiral \cite{TMMEO}). And that intractability remains when
the geometric-phase concept is extended to plane-wave scattering by 3D objects. As becomes obvious in Sec.~\ref{sec:results},
both $\Psis\sca(\ur)$ and $\Psia\sca(\ur)$ provide distinct information, so that both could be useful.

\section{Plane-Wave Scattering by a Sphere}\label{sec:results}
In order to provide illustrative results for the geometric-phase portrayal of electromagnetic scattering, let us consider 
$\Vint$ to be a sphere of radius $a$ (and so we set $\Va=\Vint$) occupied by 
a homogeneous, linear, and isotropic material. With $\Eint(\#r)$ and $\Hint(\#r)$ denoting the electric and
magnetic field phasors, respectively, induced at $\#r\in\Vint$, we focus on  five  
scattering problems. 
For all of these problems, the scattered field coefficients can be written in terms of the incident field
coefficients as  \cite{LM2016,LVVao,Beltrami}
\begin{equation}
\left.\begin{array}{l}
\label{eq28}
A\smn\three = \an \, A\smn\one + \bn \, B\smn\one\\[5pt]
B\smn\three=\cn\, A\smn\one+\dn\,B\smn\one
\end{array}\right\}\,,
\end{equation}
wherein $A\smn\one$ and $B\smn\one$ are given by Eqs.~(\ref{A1-plw}) and (\ref{B1-plw}), respectively.
The coefficients
 $\an$, $\bn$, $\cn$, and  $\dn$  appearing in
 Eqs.~(\ref{eq28})
 depend on the problem.
 
 \subsection{Standard Measures of Plane-Wave Scattering}\label{sec:mpws2}
 The various efficiencies provided in Sec.~\ref{sec:mpws1} simplify when the 3D object is a  sphere made of
 a homogeneous isotropic material. The following expressions involve the polarization state of the incident plane
 wave (through $\as$ and $\ap$) and the coefficients  $\an$, $\bn$, $\cn$, and  $\dn$ characterizing the size
 and the composition of the sphere:
\begin{eqnarray}
\nonumber
Q\sca
&=&\frac{2}{\ko^2a^2}    \sum_{n=1}^{\infty} \Bigg\{ (2n+1)\Bigg[  \vert\an\vert^2+\vert\bn\vert^2+\vert\cn\vert^2+\vert\dn\vert^2
\\[6pt]
&&
 -4\, {\rm Re} \left(\an\bn^\ast+\cn\dn^\ast\right)  \frac{{\rm Im}\left(\as\ap^\ast\right)}
 {\vert\as\vert^2+\vert\ap\vert^2}\Bigg]\Bigg\}\,,
\label{Qsca-sphere}
 \end{eqnarray}
\begin{equation}
 Q\ext=-\frac{2}{\ko^2a^2}  {\mbox Re}\left( \sum_{n=1}^{\infty} 
 \lec(2n+1)\les
\an+\dn -2(\bn+\cn)\frac{{\rm Im}\left(\as\ap^\ast\right)}
 {\vert\as\vert^2+\vert\ap\vert^2}
  \ris\ric
 \right)\,,
 \label{Qext-sphere}
 \end{equation}
 \begin{eqnarray}
 \nonumber
\Qf&=&\frac{1}{\ko^2a^2} \Bigg(
 \displaystyle{ 
 \frac{\Big\vert \sum_{n=1}^{\infty}\lec (2n+1)\les{i(\an+\dn)\as+(\bn+\cn)\ap}\ris\ric\Big\vert^2}{\vert\as\vert^2+\vert\ap\vert^2}
}
   \\&&+
  \displaystyle{ 
  \frac{\Big\vert \sum_{n=1}^{\infty}\lec (2n+1)\les{i(\an+\dn)\ap-(\bn+\cn)\as}\ris\ric\Big\vert^2}{\vert\as\vert^2+\vert\ap\vert^2}
}\Bigg)\,,
\label{Qf-sphere}
\end{eqnarray}
and
 \begin{eqnarray}
 \nonumber
 \Qb&=&\frac{1}{\ko^2a^2} \Bigg(
 \displaystyle{ 
 \frac{\Big\vert \sum_{n=1}^{\infty}\lec (-)^n(2n+1)\les{i(\an-\dn)\as+(\bn-\cn)\ap}\ris\ric\Big\vert^2}{\vert\as\vert^2+\vert\ap\vert^2}
}
\\&&+
  \displaystyle{ 
  \frac{\Big\vert \sum_{n=1}^{\infty}\lec (-)^n(2n+1)\les{i(\an-\dn)\ap-(\bn-\cn)\as}\ris\ric\Big\vert^2}{\vert\as\vert^2+\vert\ap\vert^2}
}\Bigg)\,.
\label{Qb-sphere}
\end{eqnarray}
 
 Values of $Q\ext$, $Q\sca$, $\Qf$, and $\Qb$, along with plots of  $\QD(\theta,\phi)$, $\Psis\sca(\theta,\phi)$, and $\Psia\sca(\theta,\phi)$,
 calculated for a sphere of size parameter $\koa=5$ and illuminated by
 \begin{itemize}
 \item[(a)] $s$-polarized,
 \item[(b)] $p$-polarized,
 \item[(c)] RCP, and
 \item[(d)] LCP
\end{itemize}
plane waves 
 are presented in Secs.~\ref{sec:dms}--\ref{sec:chi} for five different sets of constitutive properties and boundary conditions prevailing
 on the surface of sphere. Note that $\Psia\sca(\ur)\equiv 0$ when the incident   plane wave is RCP, arising primarily from the definition of the asymmetric
spinor ${\les{\psia\sca}(\ur)\ris}$ \cite{LakhJOSAB}.

\subsection{Dielectric-Magnetic Sphere with Charge-Free Surface}\label{sec:dms}
This is the standard problem of a dielectric-magnetic sphere with a charge-free
surface \cite{Logan, Mie1908}. The relative permittivity of the sphere material is denoted
by $\epsr$, the relative permeability by $\mur$,  the refractive index by $\nr=\sqrt{\mur}\sqrt{\epsr}$,
and the relative impedance by $\etar=\sqrt{\mur}/ \sqrt{\epsr}$.
The applicable boundary conditions  
\begin{equation}
\left.\begin{array}{l}
\ur\times\les\Einc(\#r)+\Esca(\#r)
-\Eint(\#r)\ris=\#0
\\[4pt]
\ur\times\les\Hinc(\#r)+\Hsca(\#r)-
\Hint(\#r)\ris =\#0
\end{array}\right\}\,, \quad
{r=a}\,,
\label{bc-std}
\end{equation}
deliver  
\begin{equation}
\left.\begin{array}{l}
\an = -{\gnone}/{\gntwo}
\\[5pt]
\bn=\cn= 0
\\[5pt]
\dn = -{\gnthree}/{\gnfour}
\end{array}\right\}\,,
\label{cde-std}
\end{equation}
where
\begin{subequations}
\begin{eqnarray}
{\gnone}&=&\mur\,j_n({\koa\nr})\,\psi_n\one(\koa)
-j_n(\koa)\,\psi_n\one({\koa\nr})\,,
\label{sn-def}
\\[5pt]
{\gntwo}&=&\mur\,j_n({\koa\nr})\,\psi_n\three(\koa)
-h_n\one(\koa)\,\psi_n\one({\koa\nr})\,,
\label{tn-def}
\\[5pt]
{\gnthree}&=&\epsr\,j_n({\koa\nr})\,\psi_n\one(\koa)
-j_n(\koa)\,\psi_n\one({\koa\nr})\,,
\label{pn-def}
\end{eqnarray}
and
\begin{eqnarray}
{\gnfour}&=&\epsr\,j_n({\koa\nr})\,\psi_n\three(\koa)
-h_n\one(\koa)\,\psi_n\one({\koa\nr})\,.
\label{qn-def}
\end{eqnarray}
\end{subequations}

Since $\bn=\cn=0$ for a dielectric-magnetic sphere with a charge-free
surface, the right sides of Eqs.~(\ref{Qsca-sphere})--(\ref{Qb-sphere})
simplify so that $Q\sca$, $Q\ext$, $\Qf$, and $\Qb$ do not depend
on $\as$ and $\ap$. Hence, all four of these efficiencies as well
as the absorption efficiency $Q\abs=Q\ext-Q\sca$ are
independent of the  polarization state of the incident plane wave.

\begin{figure}[!ht]
\centering
\begin{subfigure}{14cm}
    \centering\includegraphics[width=0.8\textwidth]{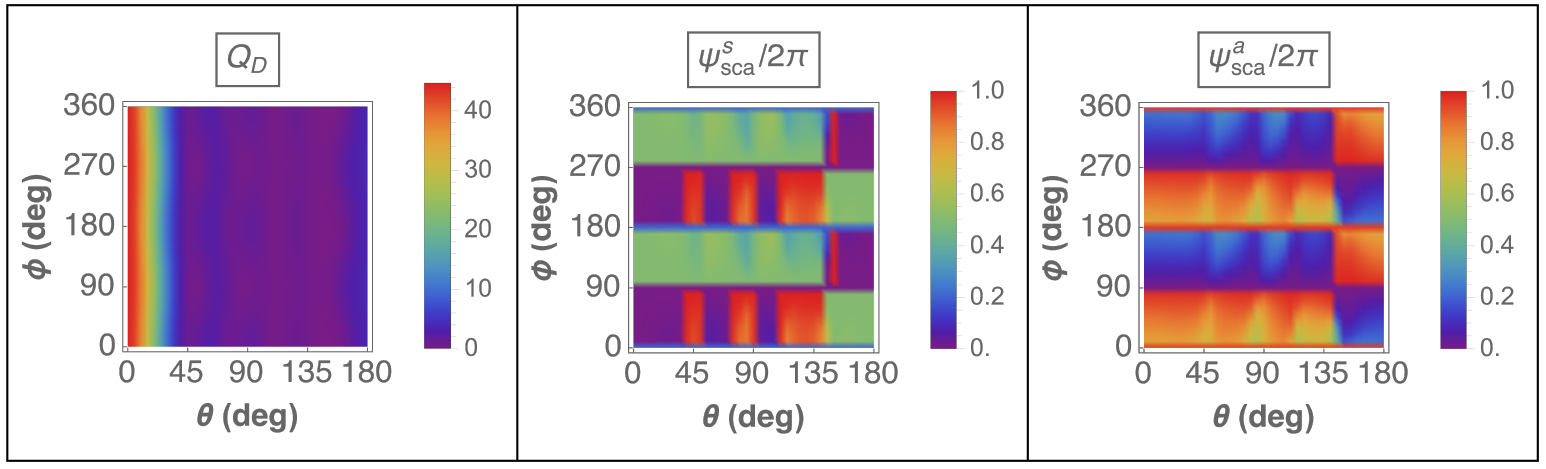}
    \caption{$s$-pol. incidence. $Q\sca=2.6637$, $Q\ext=2.6637$, $\Qf=44.735$, $\Qb=3.4704$.}
    \label{DMlossless-s}
\end{subfigure}
\hfill
\begin{subfigure}{14cm}
     \centering\includegraphics[width=0.8\textwidth]{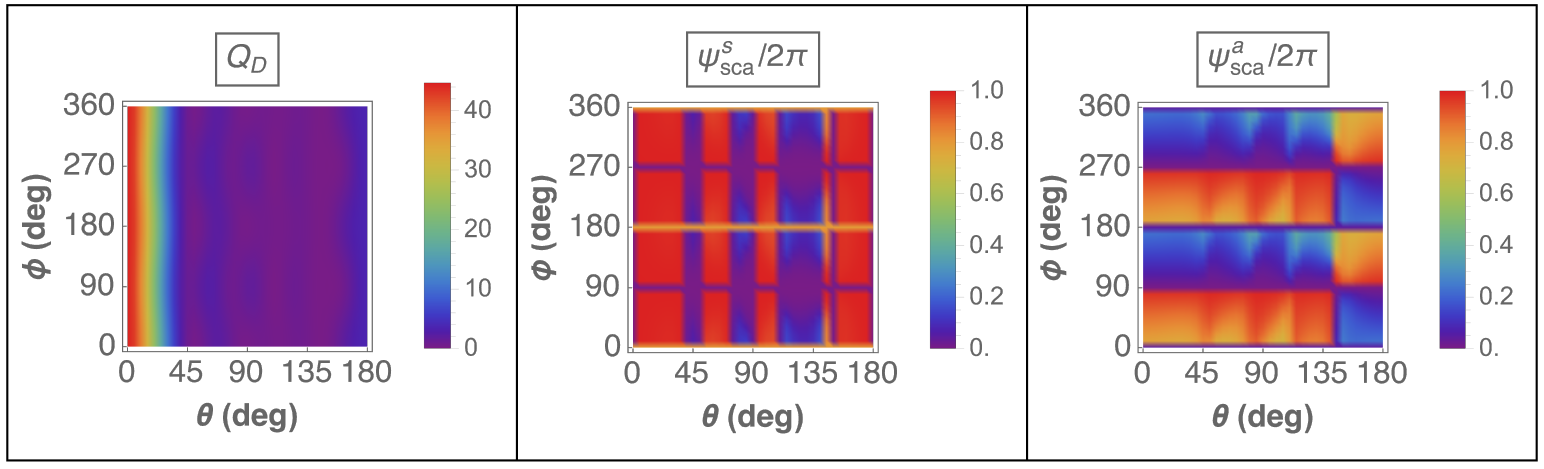}
    \caption{$p$-pol. incidence. $Q\sca=2.6637$, $Q\ext=2.6637$, $\Qf=44.735$, $\Qb=3.4704$.}
    \label{DMlossless-p}
\end{subfigure}
\hfill
\begin{subfigure}{14cm}
     \centering\includegraphics[width=0.8\textwidth]{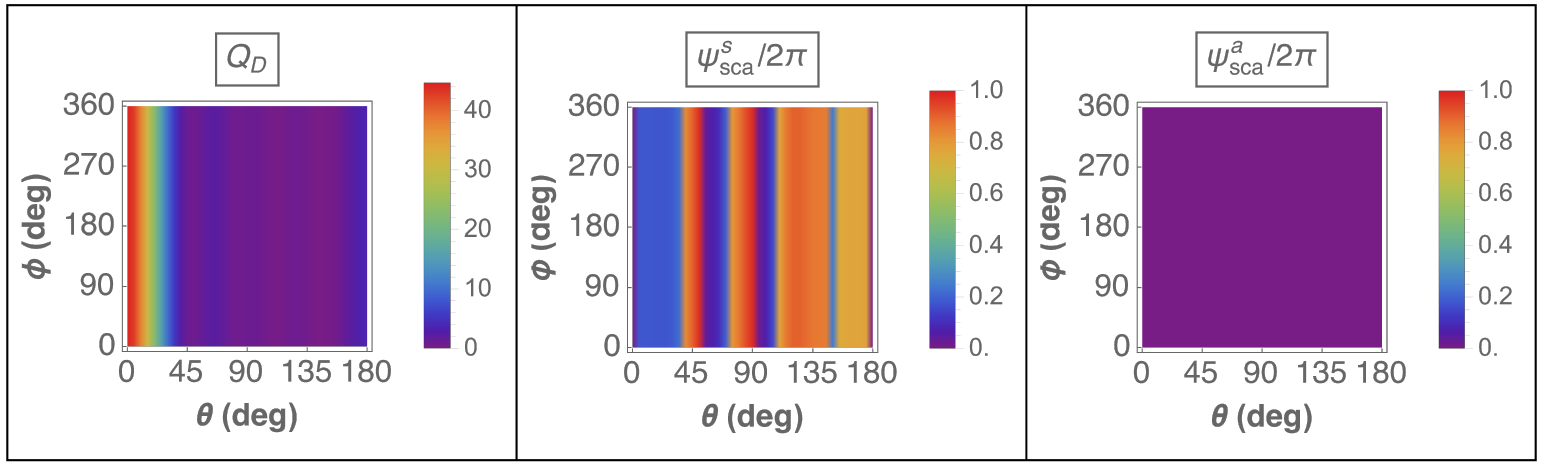}
    \caption{RCP incidence. $Q\sca=2.6637$, $Q\ext=2.6637$, $\Qf=44.735$, $\Qb=3.4704$.}
    \label{DMlossless-R}
\end{subfigure}
\hfill
\begin{subfigure}{14cm}
     \centering\includegraphics[width=0.8\textwidth]{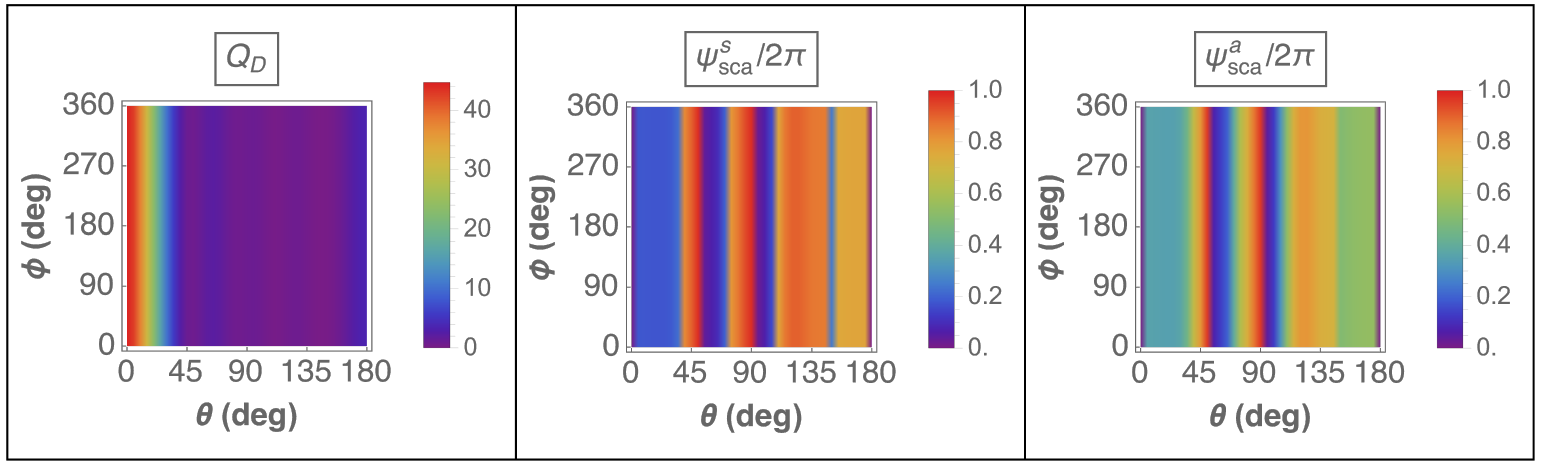}
    \caption{LCP incidence. $Q\sca=2.6637$, $Q\ext=2.6637$, $\Qf=44.735$, $\Qb=3.4704$.}
    \label{DMlossless-L}
\end{subfigure}
\caption{$\QD$, $\Psis\sca$, and $\Psia\sca$ as functions of $\theta$ and $\phi$ for
a non-dissipative dielectric-magnetic sphere with charge-free surface. The sphere
of size parameter $\koa=5$
is made of a material with $\epsr=3$ and $\mur=1.3$. }
\label{Fig:DMlossless} 
\end{figure}

Density plots of $\QD(\ur)$, $\Psis\sca(\ur)$,
and $\Psia\sca(\ur)$ as functions of the scattering angles $\theta$ and
$\phi$ are provided in Fig.~\ref{Fig:DMlossless} for a sphere of size
parameter $\koa=5$ and made of a non-dissipative dielectric-magnetic material
with $\epsr=3$ and $\mur=1.3$. As these  and the following figures were made with data calculated
at intervals $\Delta\theta=5$~deg and $\Delta\phi=5$~deg, they may not adequately
capture fine-scale features.

The following remarks can be made from Fig.~\ref{Fig:DMlossless}
and other data (not shown):
\begin{itemize}

\item \textit{Remark \ref{sec:dms}A}: $\QD(\ur)$, $\Psis\sca(\ur)$, and $\Psia\sca(\ur)$ are dependent on the polarization state
of the incident plane wave, in contrast to $Q\sca$, $Q\abs$, $Q\ext$, $\Qf$, and $\Qb$.

\item \textit{Remark \ref{sec:dms}B}:  As a function of the ratio $\as/\ap\in\mathbb{C}$, $\QD(\ur)$ for a fixed $\ur$ is: (i) either the maximum or
the minimum for the $s$-polarized incident plane wave, (ii) either the minimum or the maximum 
for the $p$-polarized incident plane waves, and (iii) is identical for left  and right  elliptically 
polarized incident plane waves with the same vibration ellipse. 

\item \textit{Remark \ref{sec:dms}C}: $\Psis\sca(\ur)$ depends on both $\theta$ and $\phi$, but only on $\theta$ when the incident plane wave is
circularly polarized.  

\item \textit{Remark \ref{sec:dms}D}: $\Psis\sca(\ur)$ is independent of the handedness of a circularly polarized incident plane wave.

\item \textit{Remark \ref{sec:dms}E}: $\Psia\sca(\ur)$ depends on both $\theta$ and $\phi$, but only on $\theta$ when the incident plane wave is
LCP.    

\item \textit{Remark \ref{sec:dms}F}: The difference $\vert\Psis\sca(\ur)-\Psia\sca(\ur)\vert$ is $\phi$-independent, when the incident plane wave is
circularly polarized.

\end{itemize}

\begin{figure}[!ht]
\centering
\begin{subfigure}{14cm}
    \centering\includegraphics[width=0.8\textwidth]{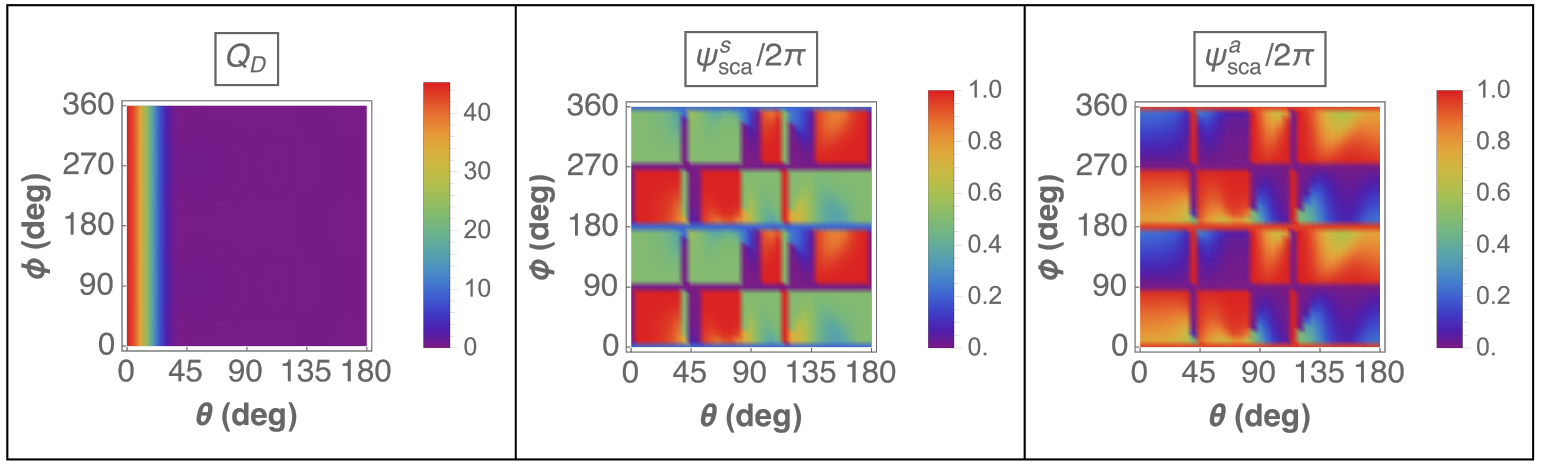}
    \caption{$s$-pol. incidence.  $Q\sca=1.2519$, $Q\ext=2.6879$, $\Qf=45.299$, $\Qb=0.0618$.}
    \label{DMlossy-s}
\end{subfigure}
\hfill
\begin{subfigure}{14cm}
     \centering\includegraphics[width=0.8\textwidth]{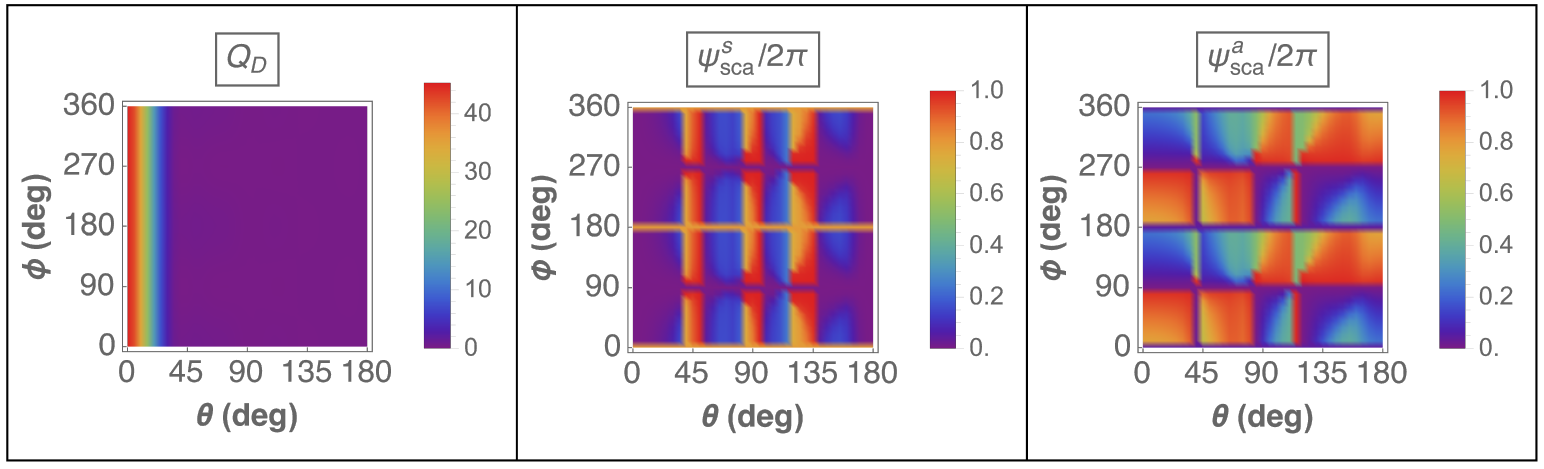}
    \caption{$p$-pol. incidence.  $Q\sca=1.2519$, $Q\ext=2.6879$, $\Qf=45.299$, $\Qb=0.0618$.}
    \label{DMlossy-p}
\end{subfigure}
\hfill
\begin{subfigure}{14cm}
     \centering\includegraphics[width=0.8\textwidth]{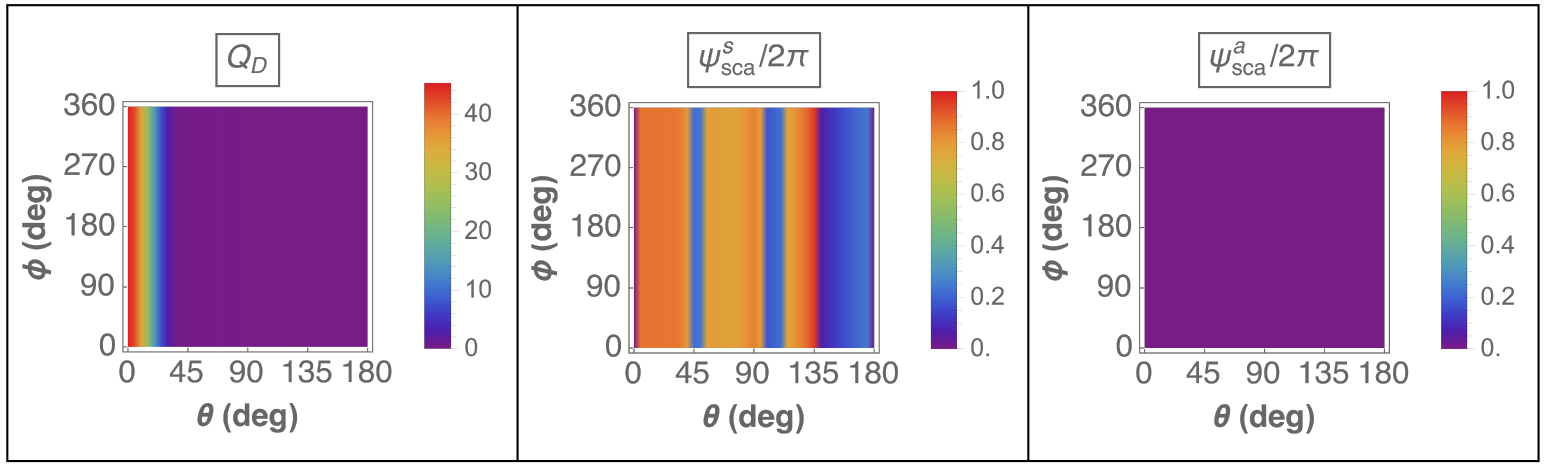}
    \caption{RCP incidence.  $Q\sca=1.2519$, $Q\ext=2.6879$, $\Qf=45.299$, $\Qb=0.0618$.}
    \label{DMlossy-R}
\end{subfigure}
\hfill
\begin{subfigure}{14cm}
     \centering\includegraphics[width=0.8\textwidth]{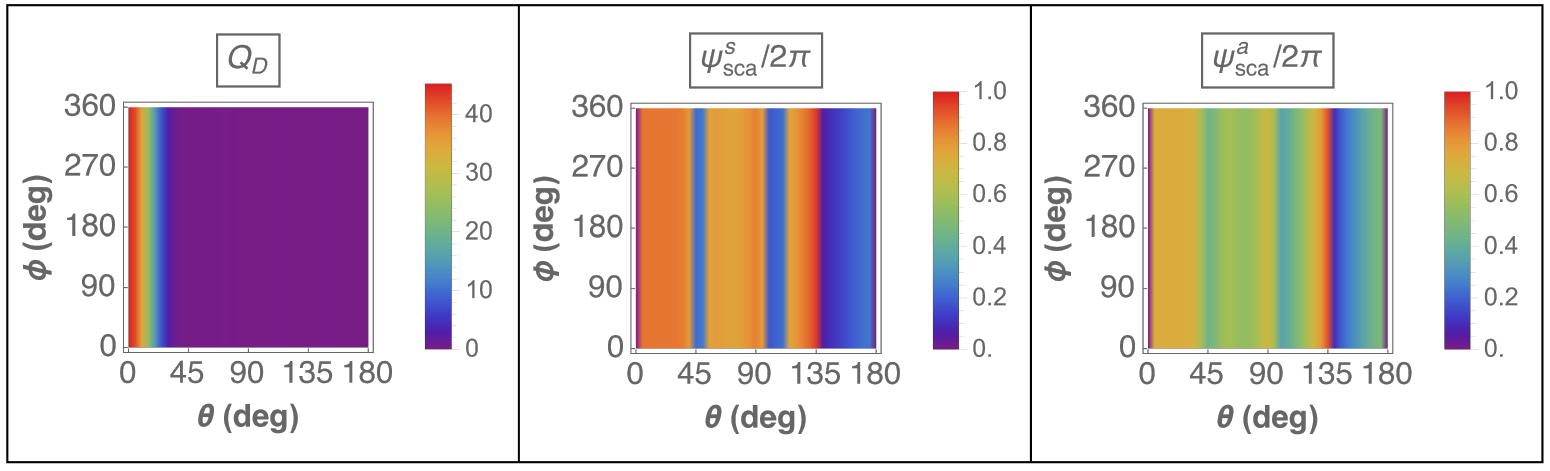}
    \caption{LCP incidence. $Q\sca=1.2519$, $Q\ext=2.6879$, $\Qf=45.299$, $\Qb=0.0618$.}
    \label{DMlossy-L}
\end{subfigure}
\caption{
Same as Fig.~\ref{Fig:DMlossless} except that
$\epsr=3(1+0.1i)$ and $\mur=1.3(1+0.1i)$. }

\label{Fig:DMlossy} 
\end{figure}

\begin{figure}[!ht]
\centering
\begin{subfigure}{14cm}
    \centering\includegraphics[width=0.8\textwidth]{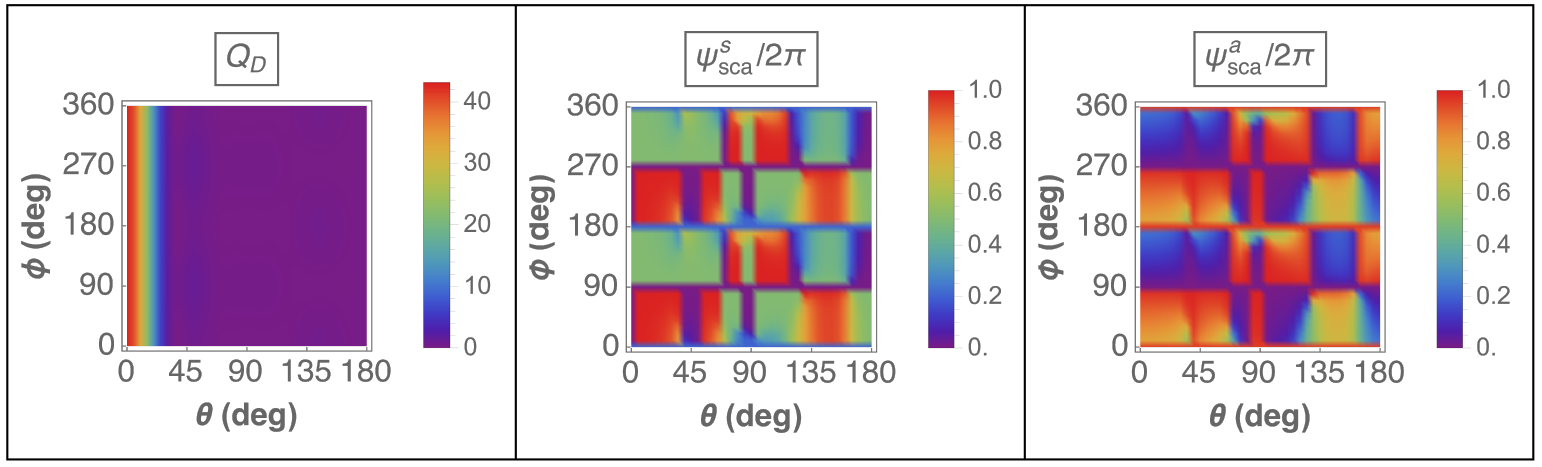}
    \caption{$s$-pol. incidence.   $Q\sca=1.2167$, $Q\ext=2.6114$, $\Qf=43.247$, $\Qb=0.0153$.}
    \label{DM-NPV-s}
\end{subfigure}
\hfill
\begin{subfigure}{14cm}
     \centering\includegraphics[width=0.8\textwidth]{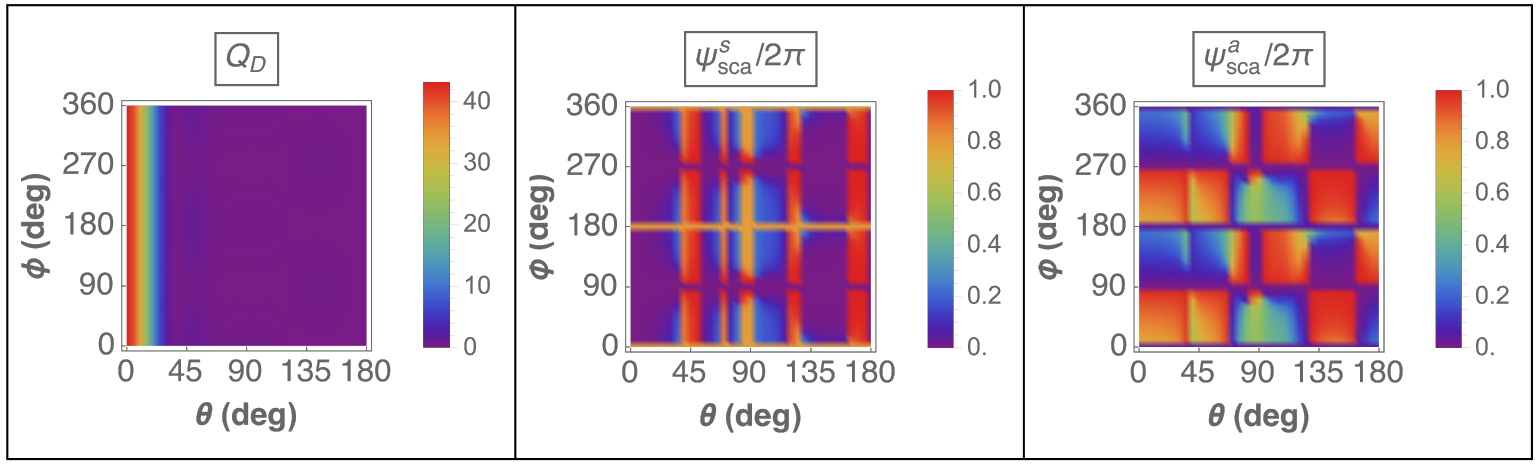}
    \caption{$p$-pol. incidence.   $Q\sca=1.2167$, $Q\ext=2.6114$, $\Qf=43.247$, $\Qb=0.0153$.}
    \label{DM-NPV-p}
\end{subfigure}
\hfill
\begin{subfigure}{14cm}
     \centering\includegraphics[width=0.8\textwidth]{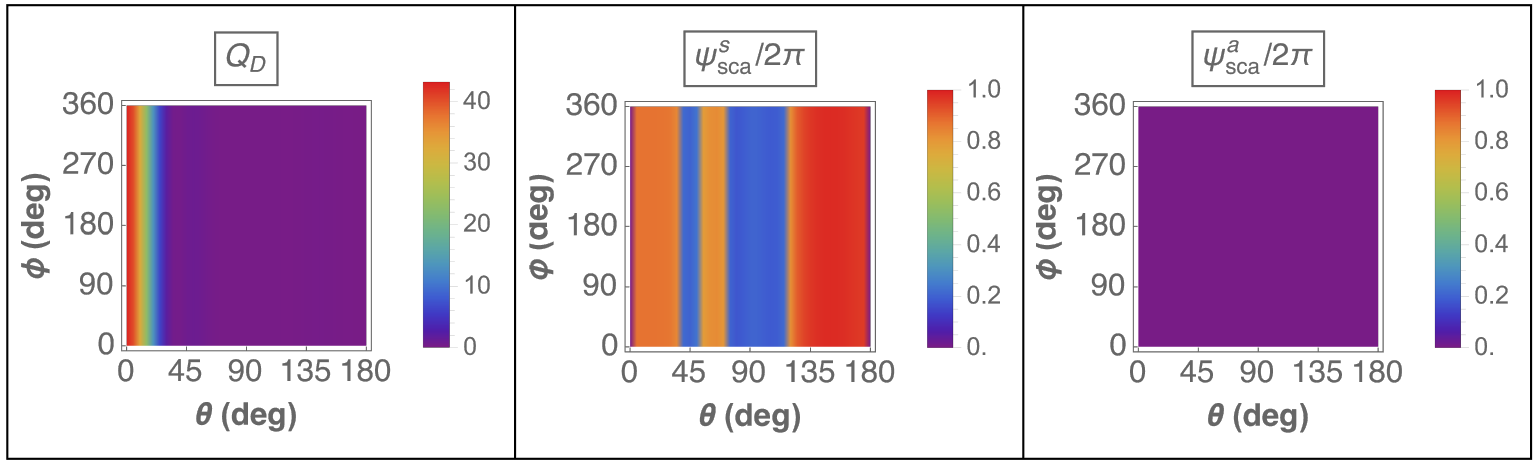}
    \caption{RCP incidence.   $Q\sca=1.2167$, $Q\ext=2.6114$, $\Qf=43.247$, $\Qb=0.0153$.}
    \label{DM-NPV-R}
\end{subfigure}
\hfill
\begin{subfigure}{14cm}
     \centering\includegraphics[width=0.8\textwidth]{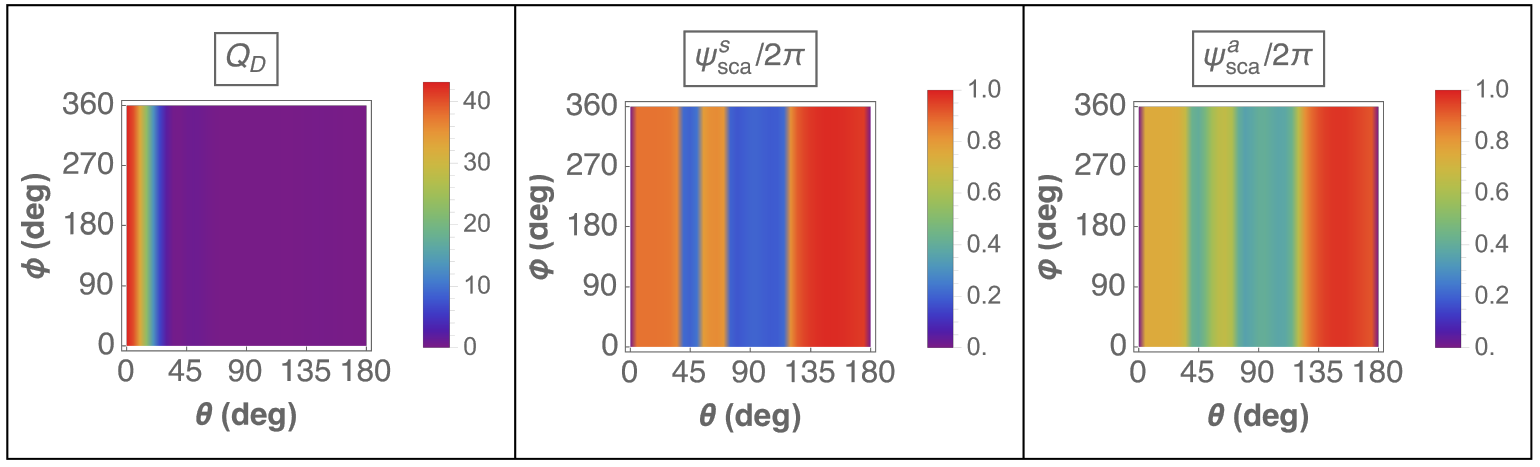}
    \caption{LCP incidence. $Q\sca=1.2167$, $Q\ext=2.6114$, $\Qf=43.247$, $\Qb=0.0153$.}
    \label{DM-NPV-L}
\end{subfigure}
\caption{
Same as Fig.~\ref{Fig:DMlossless} except that $\epsr=3(-1+0.1i)$ and $\mur=1.3(-1+0.1i)$. }
\label{Fig:DM-NPV} 
\end{figure}

All six of the foregoing remarks also apply when the sphere is made of a dissipative dielectric-magnetic material. This becomes clear from
the 
density plots of $\QD(\ur)$, $\Psis\sca(\ur)$,
and $\Psia\sca(\ur)$ as functions of the scattering angles $\theta$ and
$\phi$ provided in Fig.~\ref{Fig:DMlossy} for $\epsr=3(1+0.1i)$ and $\mur=1.3(1+0.1i)$ as
well as in Fig.~\ref{Fig:DM-NPV} for $\epsr=3(-1+0.1i)$ and $\mur=1.3(-1+0.1i)$.
Whereas the   material chosen for Fig.~\ref{Fig:DMlossy} allows positive-phase-velocity plane-wave
propagation, the   material chosen for Fig.~\ref{Fig:DM-NPV} allows negative-phase-velocity plane-wave
propagation \cite{DL}.

A perusal of literature quickly demonstrates that $Q\sca$, $Q\abs$, $Q\ext$, $\Qf$,   $\Qb$, and $\QD(\ur)$ depend on the relative permittivity and the relative permeability of the material that the sphere is made of \cite{Kerker,VMVL,BSU,vdH,BH83,MTL}, as is also
evident from the data supplied in Figs.~\ref{Fig:DMlossless}--\ref{Fig:DM-NPV}.  Furthermore, all of these quantities  depend on the
size parameter $\koa$ and therefore can be useful to address inverse problems \cite{JonesAR,CPIW,Xu,Guo,Pan,CK,Xudong,GVA}.

What, however, is remarkable is that the density plots of $\Psis\sca(\ur)$
and $\Psia\sca(\ur)$ in Figs.~\ref{Fig:DMlossless}--\ref{Fig:DM-NPV}  are far richer in identifiable features than the density plots
of $\QD(\ur)$ are. Whereas the density plots of $\QD(\ur)$ are weakly indicative of the polarization state of the incident plane wave
as well as the constitutive parameters of the sphere material in these figures, those of $\Psis\sca(\ur)$
and $\Psia\sca(\ur)$ change significantly when that polarization state and/or those constitutive parameters do;
the density plot of $\Psia\sca(\ur)$ for incident  RCP plane wave is the solitary exception.

\subsection{Impedance Sphere}  \label{sec:impsph}
The constitutive parameters of the material occupying $\Vint$ are not needed when
the impedance boundary condition
\cite{Garbacz1964,Medgyesi1985}
\begin{equation}
\ur\times\les\Einc(\#r)+\Esca(\#r)\ris= 
-\etao\etas\left(\=I-\ur\ur\right)\.
\les\Hinc(\#r)+\Hsca(\#r)\ris\,,\quad
{r=a}\,,
\label{bc-impsph}
\end{equation}
  is taken to prevail on   $\cal S$ with relative impedance $\etas$.
 As a result, the following expressions are obtained  for use in Eqs.~(\ref{eq28}):
\begin{equation}
\left.\begin{array}{l}
\an = \displaystyle{- \frac
{i\koa\, j_n(\koa)  + \etas\,\psi_n\one(\koa)}
{i\koa \,h_n\one(\koa)  + \etas \,\psi_n\three(\koa)}}
\\[5pt]
\bn=\cn=0
\\[5pt]
\dn = \displaystyle{-\frac
{i \psi_n\one(\koa) -\koa\,\etas j_n(\koa)}
{i \psi_n\three(\koa) -\koa\,\etas h_n\one(\koa)}}
\end{array}\right\}\,.
\label{cde-impsph}
\end{equation}

Furthermore, when the sphere is idealized as being made of a perfect electric conductor (PEC),
the boundary condition (\ref{bc-impsph}) simplifies to
\begin{equation}
\ur\times\les\Einc(\#r)+\Esca(\#r)\ris=\#0\,,\quad
{r=a}\,,
\label{bc-PEC}
\end{equation}
because $\etas=0$ then \cite{BSU}.  Accordingly, by setting $\etas=0$ in Eq.~(\ref{cde-impsph})
we obtain
\begin{equation}
\left.\begin{array}{l}
\an = - \displaystyle{\frac{j_n(\koa)}{h_n\one(\koa)}}
\\[5pt]
\bn=\cn=0 
\\[5pt]
\dn = -\displaystyle{\frac{\psi_n\one(\koa)  }{\psi_n\three(\koa)  }}
\end{array}\right\}\,.
\label{cde-PEC}
\end{equation}

\begin{figure}[!ht]
\centering
\begin{subfigure}{14cm}
    \centering\includegraphics[width=0.8\textwidth]{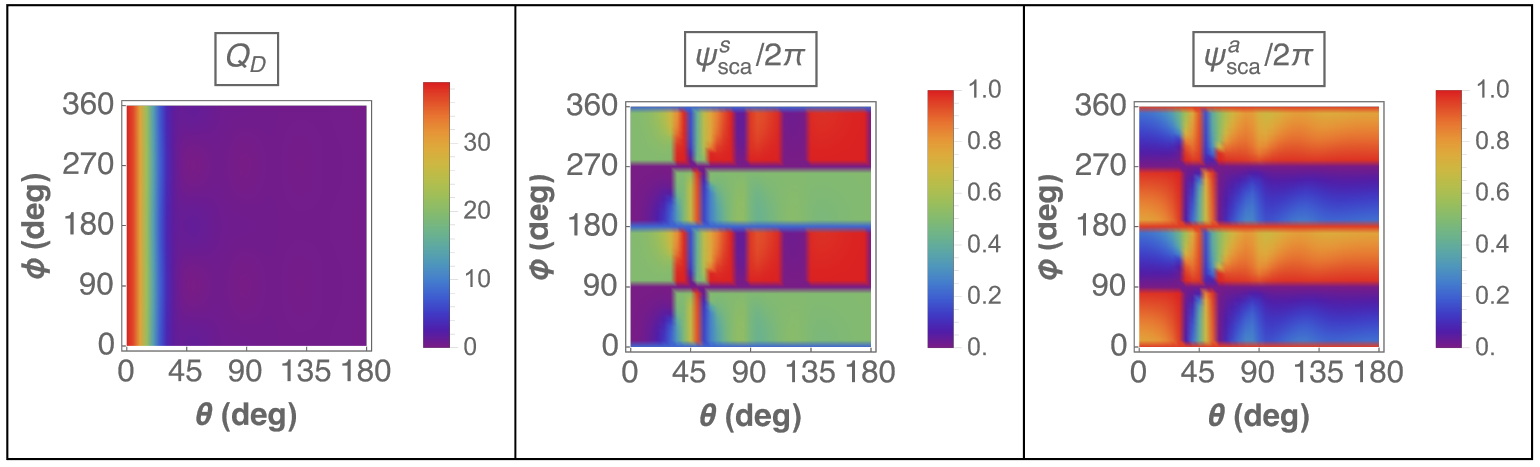}
    \caption{$s$-pol. incidence. $Q\sca=1.5011$, $Q\ext=2.4893$, $\Qf=38.985$, $\Qb=0.3168$.}
    \label{Impedance-s}
\end{subfigure}
\hfill
\begin{subfigure}{14cm}
     \centering\includegraphics[width=0.8\textwidth]{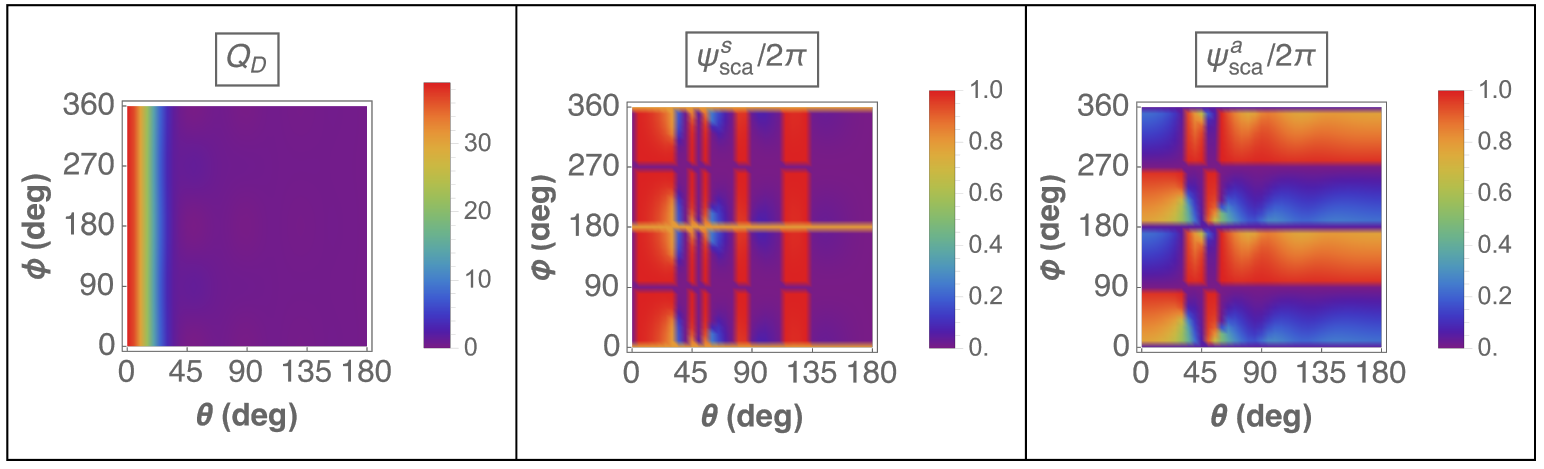}
    \caption{$p$-pol. incidence. $Q\sca=1.5011$, $Q\ext=2.4893$, $\Qf=38.985$, $\Qb=0.3168$.}
    \label{Impedance-p}
\end{subfigure}
\hfill
\begin{subfigure}{14cm}
     \centering\includegraphics[width=0.8\textwidth]{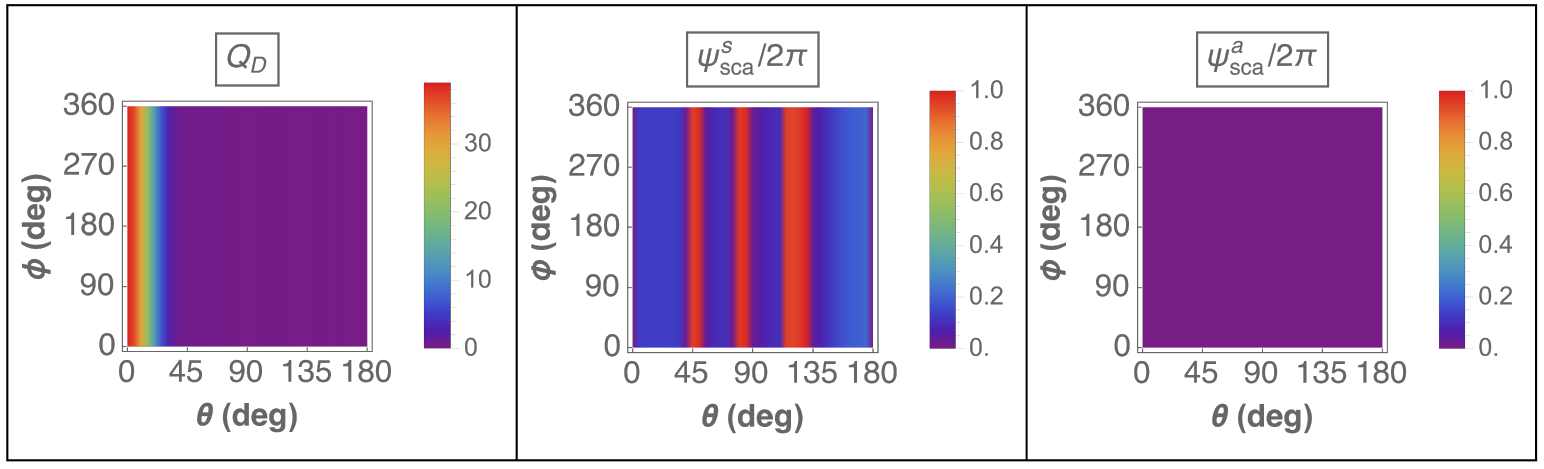}
    \caption{RCP incidence. $Q\sca=1.5011$, $Q\ext=2.4893$, $\Qf=38.985$, $\Qb=0.3168$.}
    \label{Impedance-R}
\end{subfigure}
\hfill
\begin{subfigure}{14cm}
     \centering\includegraphics[width=0.8\textwidth]{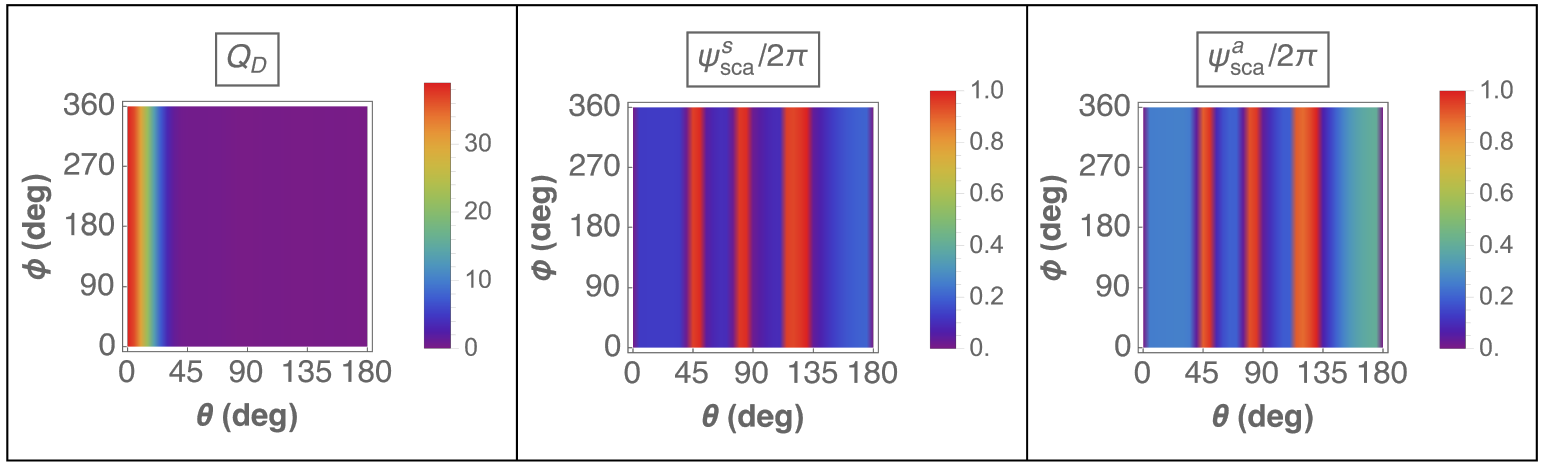}
    \caption{LCP incidence. $Q\sca=1.5011$, $Q\ext=2.4893$, $\Qf=38.985$, $\Qb=0.3168$.}
    \label{Impedance-L}
\end{subfigure}
\caption{$\QD$, $\Psis\sca$, and $\Psia\sca$ as functions of $\theta$ and $\phi$ for
an impedance sphere 
of size parameter $\koa=5$ and relative surface impedance $\etas=4$.
}
\label{Fig:Impedance} 
\end{figure}

\begin{figure}[!ht]
\centering
\begin{subfigure}{14cm}
    \centering\includegraphics[width=0.8\textwidth]{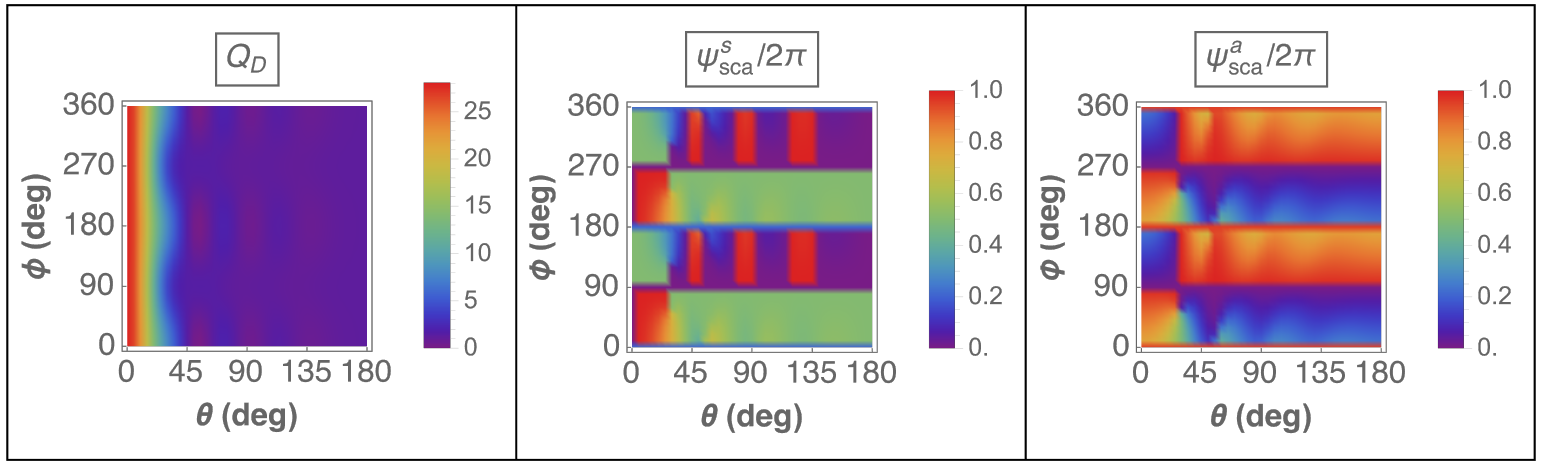}
    \caption{$s$-pol. incidence. $Q\sca=2.1161$, $Q\ext=2.1161$, $\Qf=28.073$, $\Qb=1.1688$.}
    \label{PEC-s}
\end{subfigure}
\hfill
\begin{subfigure}{14cm}
     \centering\includegraphics[width=0.8\textwidth]{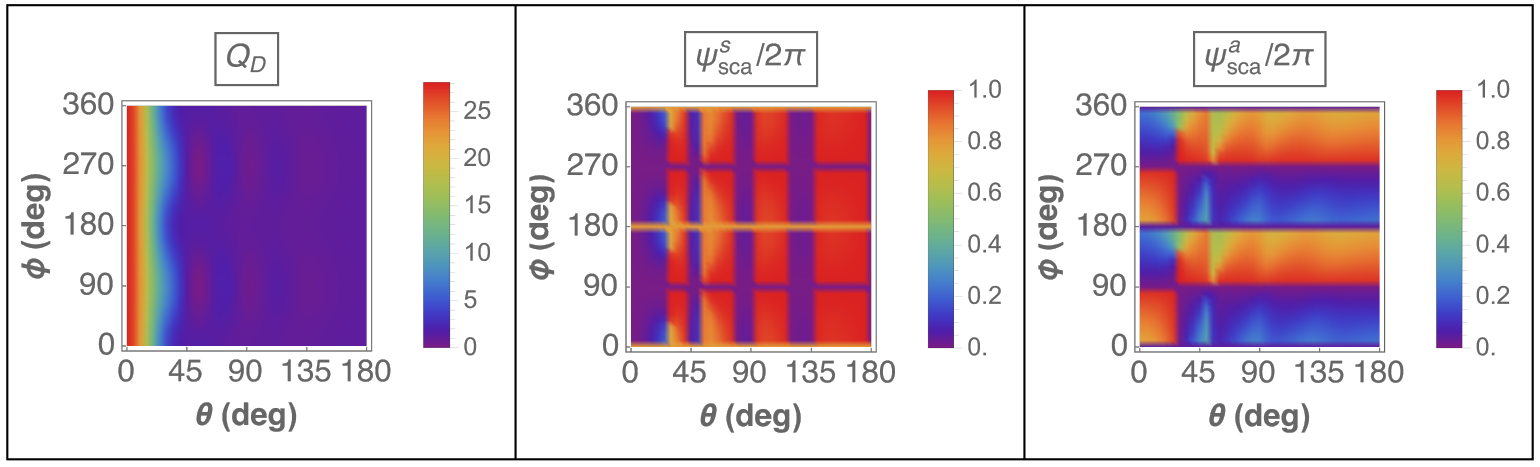}
    \caption{$p$-pol. incidence. $Q\sca=2.1161$, $Q\ext=2.1161$, $\Qf=28.073$, $\Qb=1.1688$.}
    \label{PEC-p}
\end{subfigure}
\hfill
\begin{subfigure}{14cm}
     \centering\includegraphics[width=0.8\textwidth]{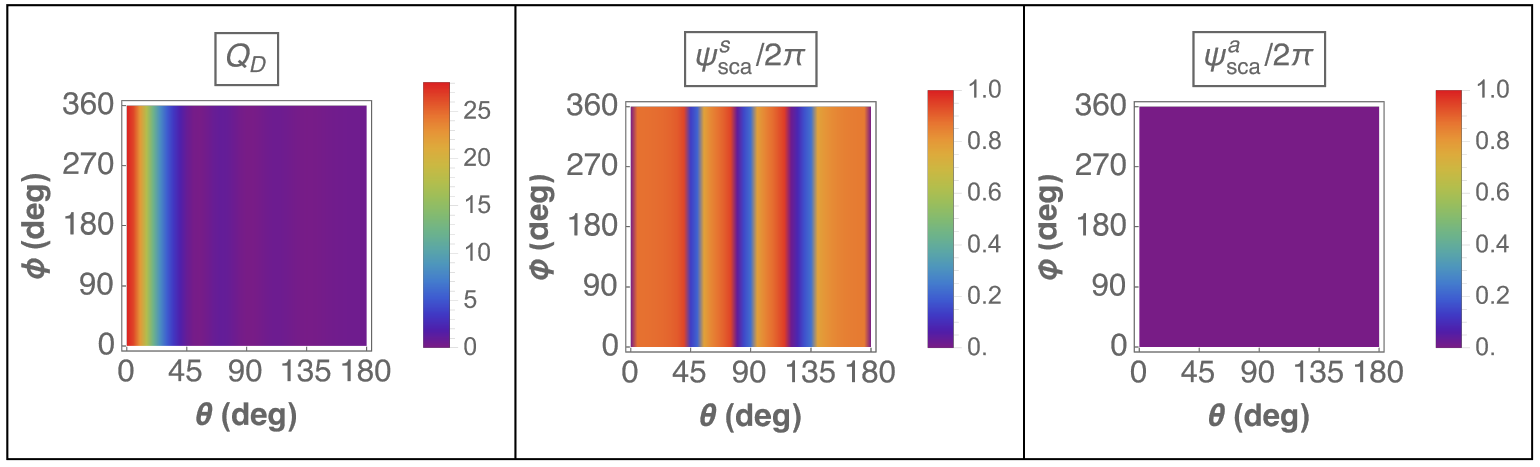}
    \caption{RCP incidence. $Q\sca=2.1161$, $Q\ext=2.1161$, $\Qf=28.073$, $\Qb=1.1688$.}
    \label{PEC-R}
\end{subfigure}
\hfill
\begin{subfigure}{14cm}
     \centering\includegraphics[width=0.8\textwidth]{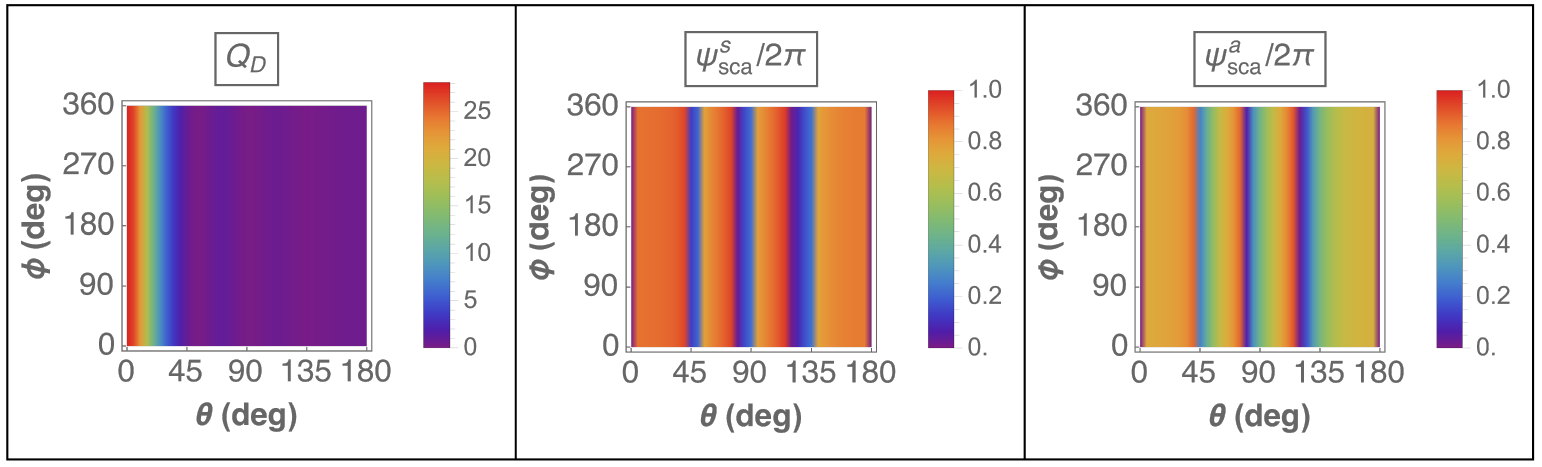}
    \caption{LCP incidence. $Q\sca=2.1161$, $Q\ext=2.1161$, $\Qf=28.073$, $\Qb=1.1688$.}
    \label{PEC-L}
\end{subfigure}
\caption{$\QD$, $\Psis\sca$, and $\Psia\sca$  as functions of $\theta$ and $\phi$ for
a perfect electrically conducting sphere 
of size parameter $\koa=5$.
}
\label{Fig:PEC} 
\end{figure}

Figures~\ref{Fig:Impedance} and \ref{Fig:PEC}, respectively, provide 
density plots of $\QD(\ur)$, $\Psis\sca(\ur)$,
and $\Psia\sca(\ur)$ for $\etas=4$ and $\etas=0$, the size
parameter of the sphere being $\koa=5$.
Similarly to the case of the dielectric-magnetic
sphere in Sec.~\ref{sec:dms},  the numerical data provided in both figures
 confirm that
 $Q\sca$, $\Qabs$, $Q\ext$, $\Qf$, and $\Qb$  of impedance and PEC spheres are
independent of the  polarization state of the incident plane wave. This is because
$\bn=\cn=0$ in Eqs.~(\ref{cde-impsph}) and (\ref{cde-PEC}). 

Remarks
\ref{sec:dms}A--F also apply to impedance and PEC spheres, as may be
concluded after examining Figs.~\ref{Fig:Impedance} and \ref{Fig:PEC}. Not only do the density plots
of $\Psis\sca(\ur)$,
and $\Psia\sca(\ur)$  clearly change when $\etas$ is changed, but the density plots of these
geometric phases are different from the ones in Figs.~\ref{Fig:DMlossless}--\ref{Fig:DM-NPV}.
That observation underscores the potential utility of geometric phases for applications of  inverse-scattering 
problems.

\subsection{Dielectric-Magnetic Sphere with Charged Surface}\label{sec:charged}
The boundary conditions (\ref{bc-std})  do not hold when a sphere made of a material with relative permittivity $\epsr$ and relative permeability $\mur$ is charged. Instead  \cite{BH1977},
\begin{equation}
\left.\begin{array}{l}
\ur\times\les\Einc(\#r)+\Esca(\#r)
-\Eint(\#r)\ris=\#0
\\[4pt]
\etao\etas\,\ur\times\les\Hinc(\#r)+\Hsca(\#r)-
\Hint(\#r)\ris= \left(\=I-\ur\ur\right)
\.\Eint(\#r)
\end{array}\right\}\,,
\quad{r=a}\,,
\label{bc-chargedsph}
\end{equation}
where  the  
relative impedance
$\etas\in\mathbb{C}$    quantifies the charge on the surface $r=a$.  
As a result \cite{BH1977,LM2016},  
\begin{equation}
\left.\begin{array}{l}
\an = \displaystyle{- \frac
{i \mur\koa\, j_n(\koa)\, j_n({\koa\nr}) + \etas\,{\gnone}}
{i \mur\koa\, h_n\one(\koa) \,j_n({\koa\nr}) + \etas\,{\gntwo}}}
\\[5pt]
\bn=\cn  =0
\\[5pt]
\dn = \displaystyle{-\frac
{i\psi_n\one(\koa)\,\psi_n\one({\koa\nr})+\koa\, \etas\,{\gnthree}}
{i\psi_n\three(\koa)\,\psi_n\one({\koa\nr})+\koa\,\etas\, {\gnfour}}}
\end{array}\right\}\,.
\label{cde-chargedsph}
\end{equation}
Note that $\etas\in\mathbb{R}$ at low frequencies \cite{BH1977}.

\begin{figure}[!ht]
\centering
\begin{subfigure}{14cm}
    \centering\includegraphics[width=0.8\textwidth]{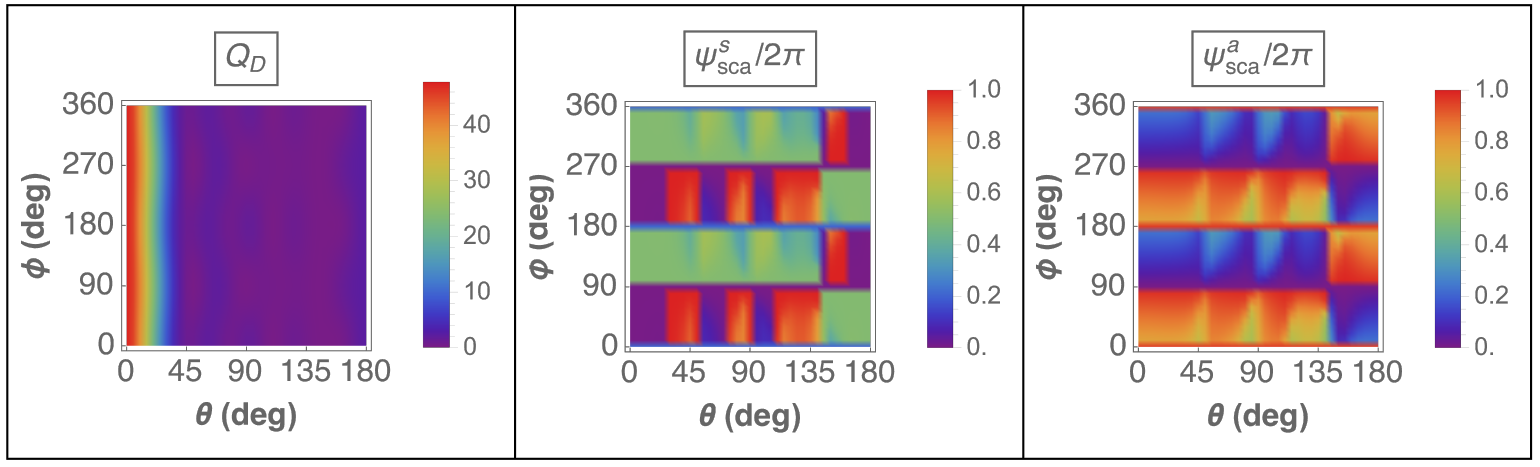}
    \caption{$s$-pol. incidence. $Q\sca=2.4920$, $Q\ext=2.7565$, $\Qf=47.797$, $\Qb=2.2521$.}
    \label{Charged-s}
\end{subfigure}
\hfill
\begin{subfigure}{14cm}
     \centering\includegraphics[width=0.8\textwidth]{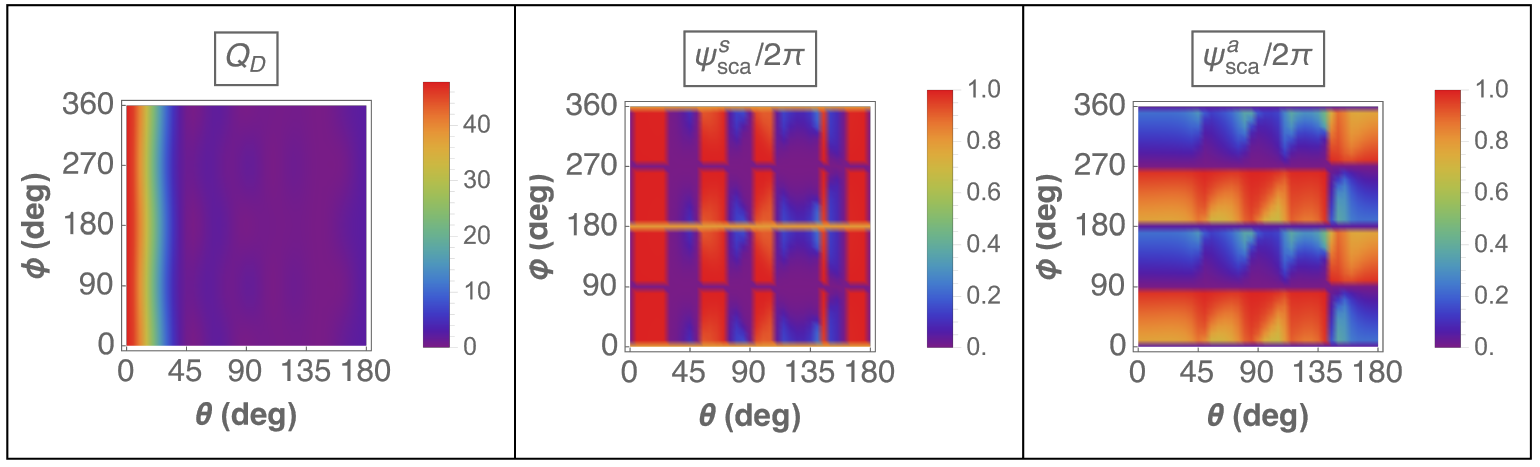}
    \caption{$p$-pol. incidence. $Q\sca=2.4920$, $Q\ext=2.7565$, $\Qf=47.797$, $\Qb=2.2521$.}
    \label{Charged-p}
\end{subfigure}
\hfill
\begin{subfigure}{14cm}
     \centering\includegraphics[width=0.8\textwidth]{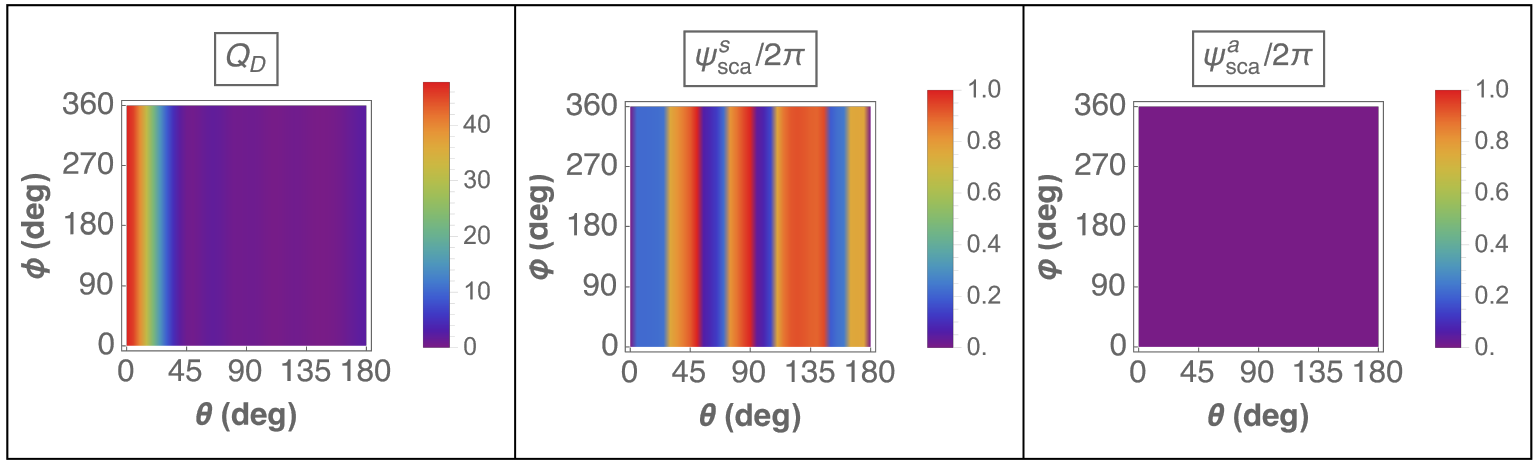}
    \caption{RCP incidence. $Q\sca=2.4920$, $Q\ext=2.7565$, $\Qf=47.797$, $\Qb=2.2521$.}
    \label{Charged-R}
\end{subfigure}
\hfill
\begin{subfigure}{14cm}
     \centering\includegraphics[width=0.8\textwidth]{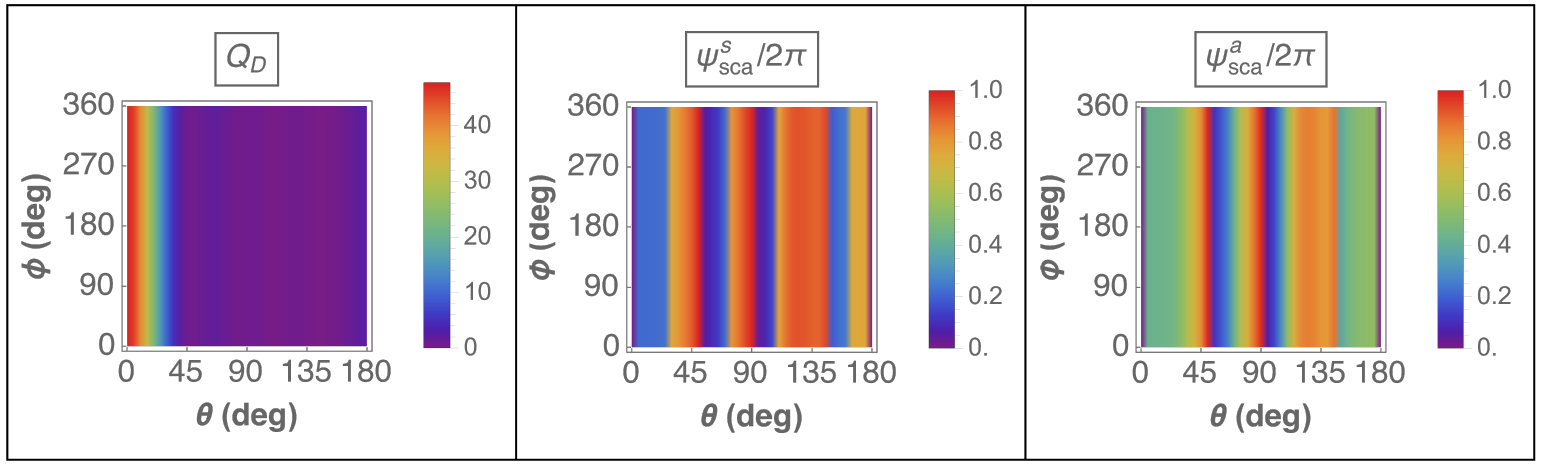}
    \caption{LCP incidence. $Q\sca=2.4920$, $Q\ext=2.7565$, $\Qf=47.797$, $\Qb=2.2521$.}
    \label{Charged-L}
\end{subfigure}
\caption{$\QD$, $\Psis\sca$, and $\Psia\sca$ as functions of $\theta$ and $\phi$ for
a non-dissipative dielectric-magnetic sphere with a charged surface. The sphere
of size parameter $\koa=5$
is made of a material with $\epsr=3$ and $\mur=1.3$. The  charge on the surface $r=a$ is quantified through the relative impedance
$\etas=10$. }
\label{Fig:Charged} 
\end{figure}

Density plots of $\QD(\ur)$, $\Psis\sca(\ur)$,
and $\Psia\sca(\ur)$ for a charged sphere characterized by
$\koa=5$,  $\epsr=3$, $\mur=1.3$, and $\etas=10$ are provided in Fig.~\ref{Fig:Charged}.
Once again,  $Q\sca$, $\Qabs$, $Q\ext$, $\Qf$, and $\Qb$  turn out to be
independent of the  polarization state of the incident plane wave, because
$\bn=\cn=0$ in Eqs.~(\ref{cde-chargedsph}). Additionally, Remarks
\ref{sec:dms}A--F still apply. A comparison of Figs.~\ref{Fig:DMlossless} and  \ref{Fig:Charged}
indicates that charging does have a small but noticeable effect on $\Psis\sca(\ur)$
and $\Psia\sca(\ur)$.

\subsection{Dielectric-Magnetic Sphere with Topologically Insulating Surface States}\label{sec:TI}
 When a sphere made of a material with relative permittivity $\epsr$ and relative permeability $\mur$  has topologically insulating surface states \cite{LM2016}, the
boundary conditions to be satisfied are
\begin{equation}
\left.\begin{array}{l}
\ur\times\les\Einc(\#r)+\Esca(\#r)
-\Eint(\#r)\ris=\#0
\\[4pt]
\ur\times\les\Hinc(\#r)+\Hsca(\#r)-
\Hint(\#r)\ris= -\gammas\ur\times\Eint(\#r)
\end{array}\right\}\,,\quad
{r=a}\,,
\label{bc-TI}
\end{equation}
where the surface admittance $\gammas\in\mathbb{R}$ is quantized in terms of $\etao^{-1}\tilde{\alpha}$,
 $\tilde{\alpha}$ being the dimensionless fine structure constant \cite{Maciejko}. The situation is different from that of a charged sphere,
 as may be appreciated by comparing Eqs.~(\ref{bc-chargedsph}) and (\ref{bc-TI}) and noting that $\=I-\ur\ur = -(\ur\times\=I)\.(\ur\times\=I)$.

Satisfaction of the boundary conditions (\ref{bc-TI}) yields \cite{LM2016}
\begin{equation}
\left.\begin{array}{l}
\an = \displaystyle{-\frac{{\gnone}\,{\gnfour}
-{\etao^2\gammas^2}\mur\,j_n(\koa)\,\psi_n\three(\koa)\,
j_n({\koa\nr})\,\psi_n\one({\koa\nr})}
{{\gntwo}\,{\gnfour}
-{\etao^2\gammas^2}\mur\,
h_n\one(\koa)\,\psi_n\three(\koa)\,
j_n({\koa\nr})\,\psi_n\one({\koa\nr})}}
\\[12pt]
\bn=-\cn=\displaystyle{ \frac{\left(\etao\gammas \mur/\koa\right)\,
j_n({\koa\nr})\,\psi_n\one({\koa\nr})}
{{\gntwo}\,{\gnfour}
-{\etao^2\gammas^2}\mur\,
h_n\one(\koa)\,\psi_n\three(\koa)\,
j_n({\koa\nr})\,\psi_n\one({\koa\nr})}}
\\[12pt]
\dn =\displaystyle{ -\frac{{\gntwo}\,{\gnthree}
-{\etao^2\gammas^2}\mur\,
h_n\one(\koa)\,\psi_n\one(\koa)\,
j_n({\koa\nr})\,\psi_n\one({\koa\nr})}
{{\gntwo}\,{\gnfour}
-{\etao^2\gammas^2} \mur\,
h_n\one(\koa)\,\psi_n\three(\koa)\,
j_n({\koa\nr})\,\psi_n\one({\koa\nr})}}
\end{array}\right\}\,.
\label{cde-TI}
\end{equation}
Thus, the replacement of $(\ur\times\=I)\.(\ur\times\=I)$ in  Eqs.~(\ref{bc-chargedsph}) by $(\ur\times\=I)$ 
leads to non-zero values
for $\bn$ and $\cn$ when the sphere has
topologically insulating surface states.

Since $\bn=-\cn\ne 0$, Eqs.~(\ref{Qext-sphere}) and (\ref{Qf-sphere})
inform us that
 $Q\ext$ and $\Qf$ are definitely independent
of the polarization state of the incident plane wave, for a sphere with topologically insulating surface states.

\begin{figure}[!ht]
\centering
\begin{subfigure}{14cm}
    \centering\includegraphics[width=0.8\textwidth]{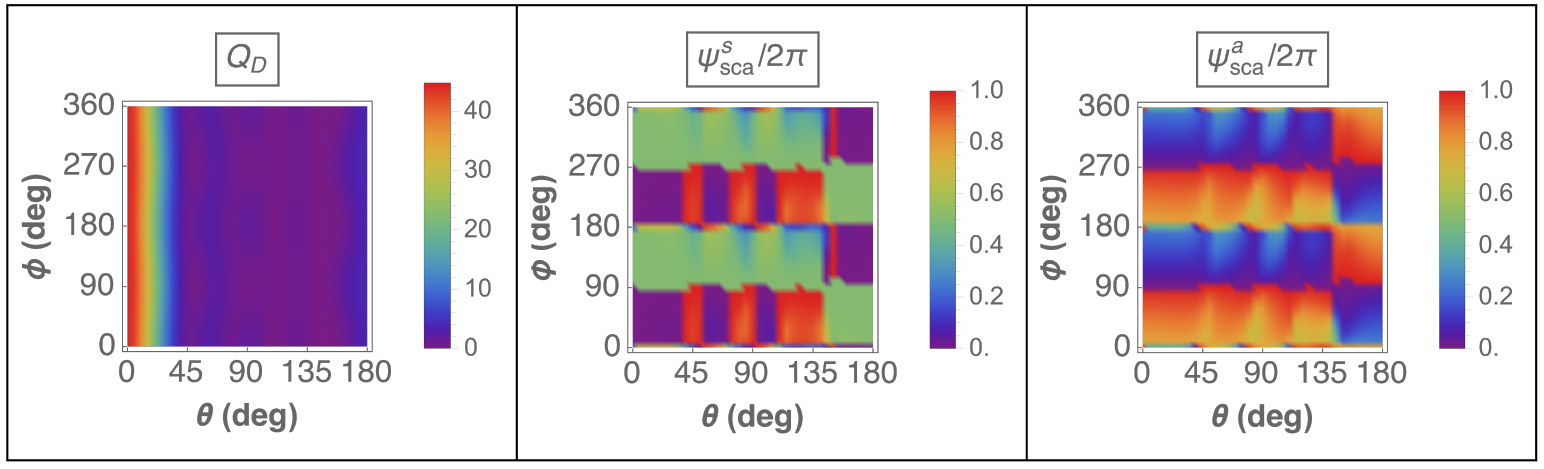}
    \caption{$s$-pol. incidence. $Q\sca=2.6647$, $Q\ext=2.6647$, $\Qf=44.735$, $\Qb=3.5444$.}
    \label{TIlossless-s}
\end{subfigure}
\hfill
\begin{subfigure}{14cm}
     \centering\includegraphics[width=0.8\textwidth]{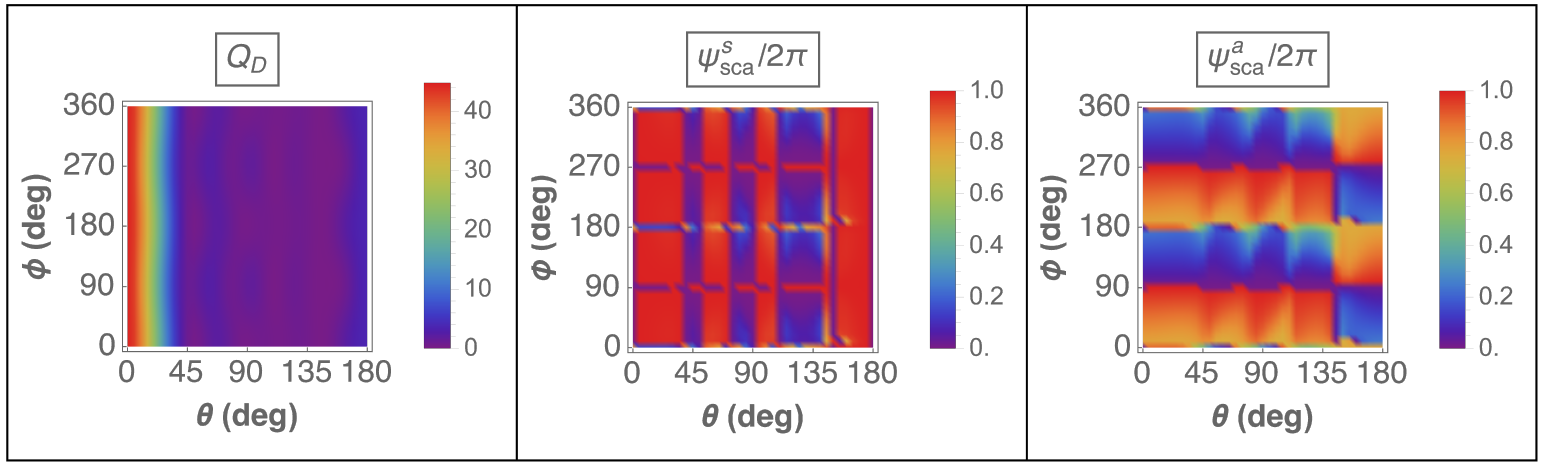}
    \caption{$p$-pol. incidence. $Q\sca=2.6647$, $Q\ext=2.6647$, $\Qf=44.735$, $\Qb=3.5444$.}
    \label{TIlossless-p}
\end{subfigure}
\hfill
\begin{subfigure}{14cm}
     \centering\includegraphics[width=0.8\textwidth]{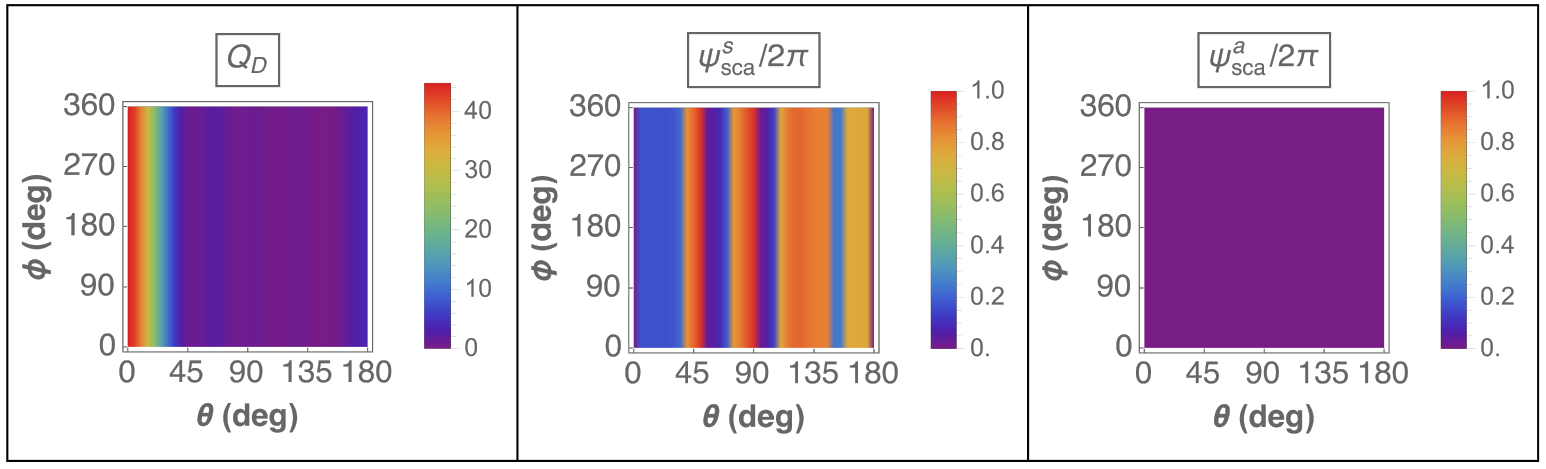}
    \caption{RCP incidence. $Q\sca=2.6647$, $Q\ext=2.6647$, $\Qf=44.735$, $\Qb=3.5444$.}
    \label{TIlossless-R}
\end{subfigure}
\hfill
\begin{subfigure}{14cm}
     \centering\includegraphics[width=0.8\textwidth]{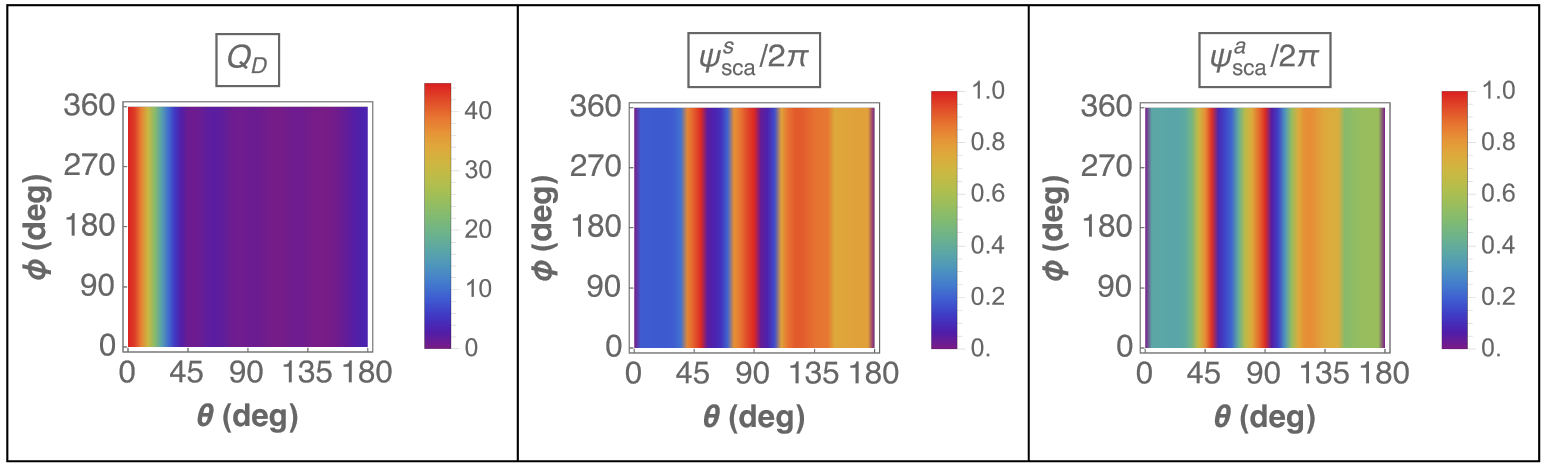}
    \caption{LCP incidence. $Q\sca=2.6647$, $Q\ext=2.6647$, $\Qf=44.735$, $\Qb=3.5444$.}
    \label{TIlossless-L}
\end{subfigure}
\caption{$\QD$, $\Psis\sca$, and $\Psia\sca$  as functions of $\theta$ and $\phi$ for
a non-dissipative dielectric-magnetic sphere with topologically insulating surface states. The sphere
of size parameter $\koa=5$
is made of a material with $\epsr=3$ and $\mur=1.3$. The surface states are characterized by the surface
admittance $\gammas=10\etao^{-1}\tilde{\alpha}$. }
\label{Fig:TIlossless} 
\end{figure}

\begin{figure}[!ht]
\centering
\begin{subfigure}{14cm}
    \centering\includegraphics[width=0.8\textwidth]{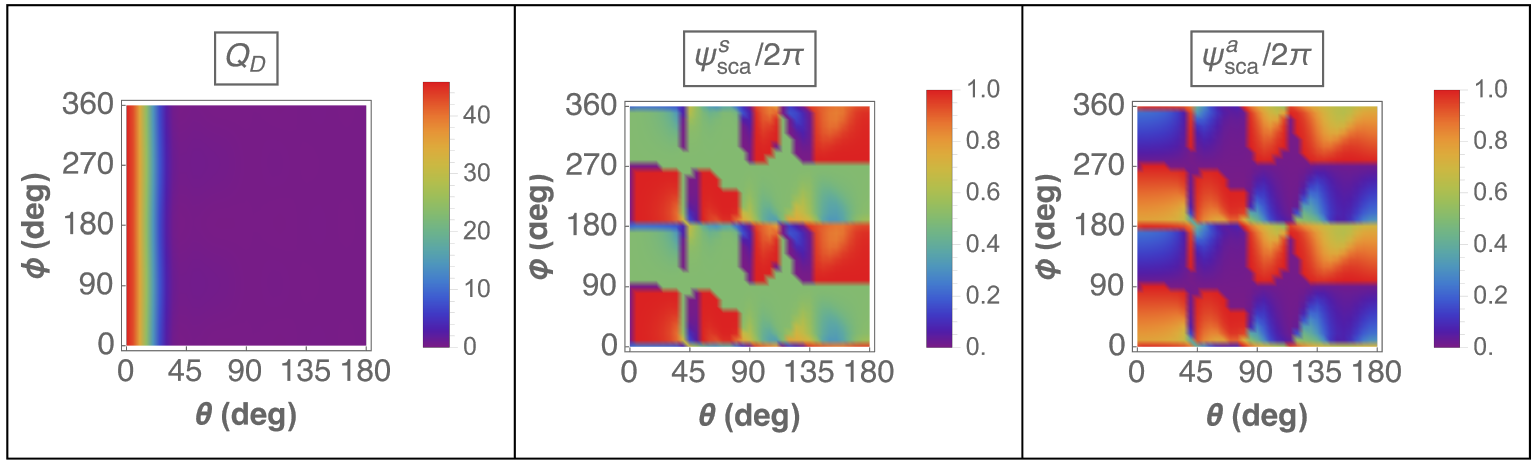}
    \caption{$s$-pol. incidence. $Q\sca=1.2728$, $Q\ext=2.7071$, $\Qf=45.935$, $\Qb=0.0707$.}
    \label{TIlossy-s}
\end{subfigure}
\hfill
\begin{subfigure}{14cm}
     \centering\includegraphics[width=0.8\textwidth]{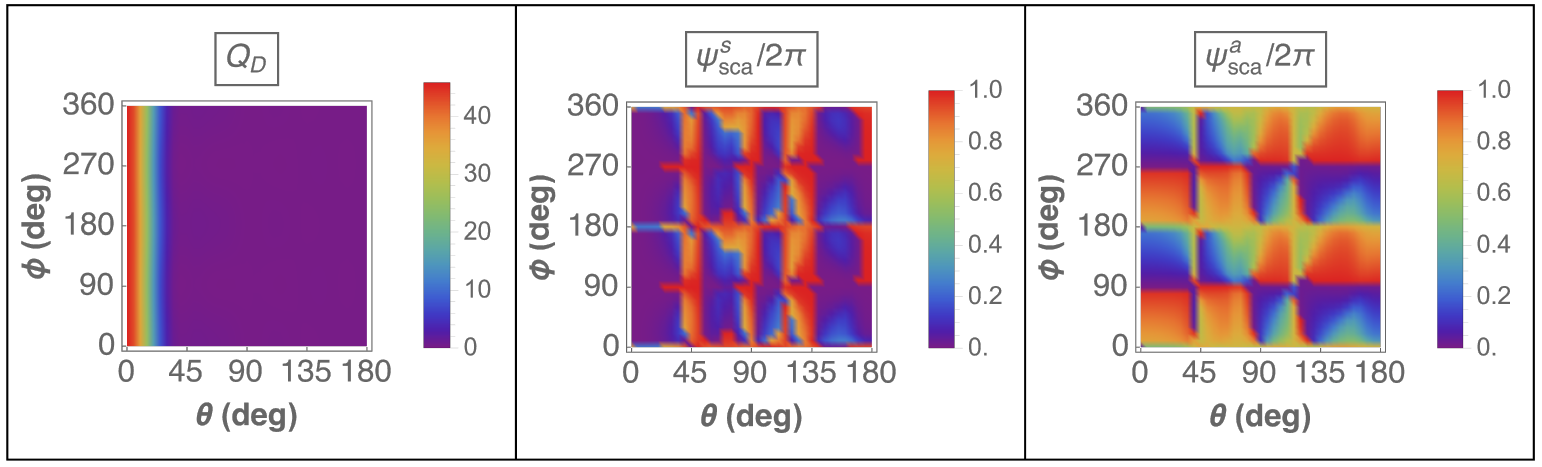}
    \caption{$p$-pol. incidence. $Q\sca=1.2728$, $Q\ext=2.7071$, $\Qf=45.935$, $\Qb=0.0707$.}
    \label{TIlossy-p}
\end{subfigure}
\hfill
\begin{subfigure}{14cm}
     \centering\includegraphics[width=0.8\textwidth]{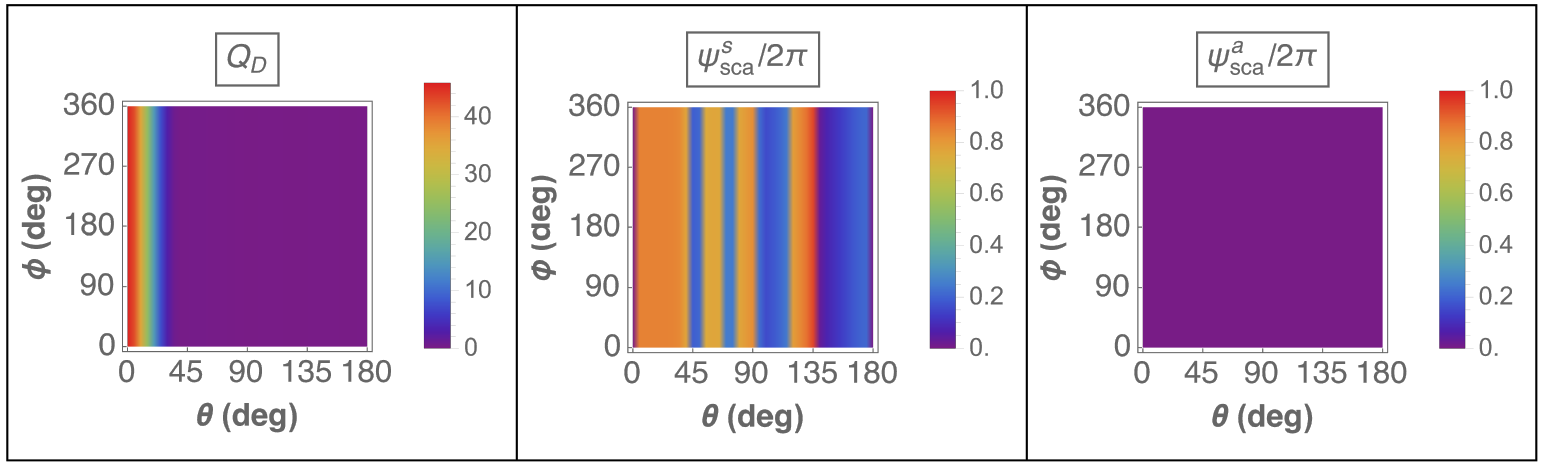}
    \caption{RCP incidence. $Q\sca=1.2687$, $Q\ext=2.7071$, $\Qf=45.935$, $\Qb=0.0658$.}
    \label{TIlossy-R}
\end{subfigure}
\hfill
\begin{subfigure}{14cm}
     \centering\includegraphics[width=0.8\textwidth]{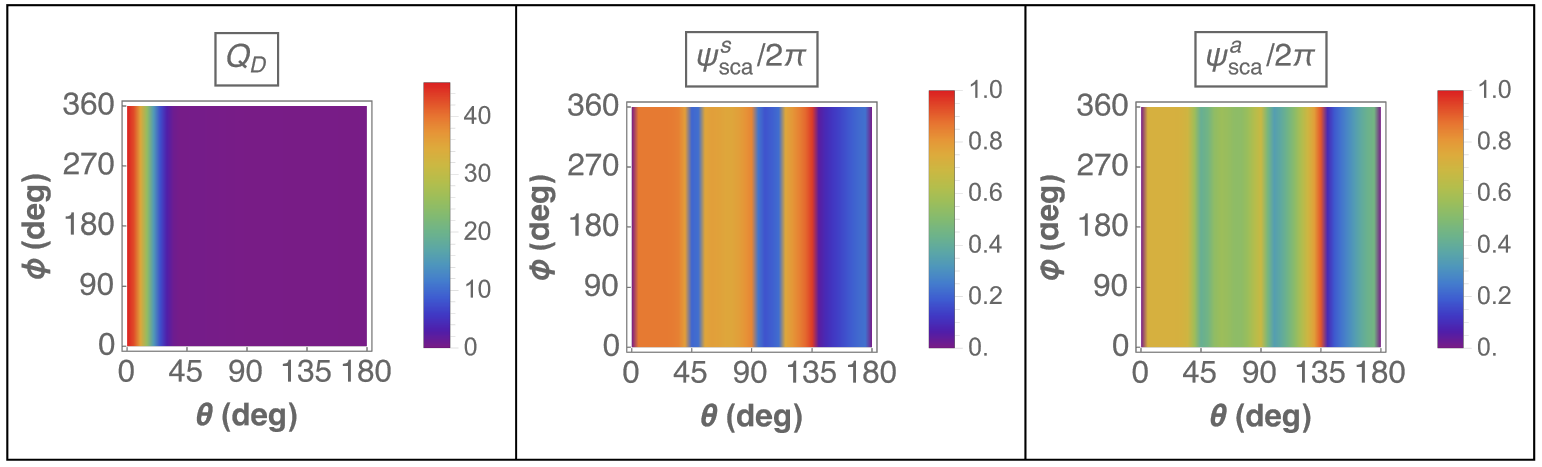}
    \caption{LCP incidence. $Q\sca=1.2770$, $Q\ext=2.7071$, $\Qf=45.935$, $\Qb=0.0755$.}
    \label{TIlossy-L}
\end{subfigure}
\caption{
Same as Fig.~\ref{Fig:TIlossless} except that $\epsr=3(1+0.2i)$. }
\label{Fig:TIlossy} 
\end{figure}

Density plots of $\QD(\ur)$, $\Psis\sca(\ur)$,
and $\Psia\sca(\ur)$ for a sphere characterized by
$\koa=5$,  $\epsr=3$, $\mur=1.3$, and $\gammas=10\etao^{-1}\tilde{\alpha}$ are provided in Fig.~\ref{Fig:TIlossless}.
Analogous density plots  are provided in Fig.~\ref{Fig:TIlossy} for the same parameters except that $\epsr=3(1+0.2i)$.
The following remarks can be made from these two figures
and other data (not shown):
\begin{itemize}

\item \textit{Remark \ref{sec:TI}A}: $\QD$, $\Psis\sca$, and $\Psia\sca$ are dependent on the polarization state
of the incident plane wave, but   $Q\ext$ and $\Qf$ are not. Additionally, $Q\sca$, $Q\abs$, and $\Qb$ are independent
of the same polarization state, provided that $\etar\in\mathbb{R}$.

\item \textit{Remark \ref{sec:TI}B}:  As a function of the ratio $\as/\ap\in\mathbb{C}$, $\QD(\ur)$ for a fixed $\ur$ is: (i) either the maximum or
the minimum for the $s$-polarized incident plane wave and (ii) either the minimum or the maximum 
for the $p$-polarized incident plane waves. Furthermore, $\QD(\ur)$ for a fixed $\ur$ is: (iii)  identical for left  and right  elliptically 
polarized incident plane waves with the same vibration ellipse, provided that $\etar\in\mathbb{R}$.

\item \textit{Remark \ref{sec:TI}C}: $\Psis\sca(\ur)$ depends on both $\theta$ and $\phi$, but only on $\theta$ when the incident plane wave is
circularly polarized.

\item \textit{Remark \ref{sec:TI}D}: $\Psia\sca(\ur)$ depends on both $\theta$ and $\phi$, but only on $\theta$ when the incident plane wave is
LCP.   

\item \textit{Remark \ref{sec:TI}E}: The difference $\vert\Psis\sca(\ur)-\Psia\sca(\ur)\vert$ is $\phi$-independent, when the incident plane wave is
circularly polarized.

\end{itemize}

 \subsection{Isotropic Chiral Sphere}\label{sec:chi}
The Tellegen constitutive equations
of an isotropic chiral material can be stated as \cite{Beltrami}
\begin{equation}
\left.
\begin{array}{l}
    \#{D} = \epsr \, \epso \, \#{E} + i \, \kappa \sqrt{\epso\muo} \, \#{H} 
    \\[5pt]
    \#{B} = \mur \, \muo \,\#{H} - i \, \kappa \sqrt{\epso \muo} \, \#{E}
\end{array}
\right\}\,,
\label{def-chiral}
\end{equation}
where $\kappa$ is the chirality parameter. Although the sphere is made of this material,
the standard boundary conditions (\ref{bc-std}) still hold, it being assumed that the
surface of the sphere is charge free \cite{Bohren1974}. The satisfaction of those boundary conditions
delivers \cite{Hector}
\begin{equation}
\left.\begin{array}{l}
\an=
\left({\fnone}{\fnthree}-2 {\etar}{\fnsix}\right)/{\fneight}
\\[5pt]
\bn=\cn=-{\etar} {\fntwo}{\fnfive}/{\fneight}
\\[5pt]
\dn=
\left({\fnone}{\fnfour}-2 {\etar} {\fnsix}\right)/{\fneight}
\end{array}
\right\}\,,
\end{equation}
where
\begin{subequations}
\begin{eqnarray}
{\fnone}&=&\displaystyle{{j_n(\kRa)} {\frac{\psi_n\one(\kLa)}{\kLa}}+{j_n(\kLa)} {\frac{\psi_n\one(\kRa)}{\kRa}}}
\\[5pt]
{\fntwo}&=&\displaystyle{{j_n(\kRa)} {\frac{\psi_n\one(\kLa)}{\kLa}}-{j_n(\kLa)} {\frac{\psi_n\one(\kRa)}{\kRa}}}
\\[5pt]
{\fnthree}&=& \displaystyle{ {\etar^2}\, {h_n\one(\koa)} {\frac{\psi_n\one(\koa)}{\koa}}+{j_n(\koa)} {\frac{\psi_n\three(\koa)}{\koa}}}
\\[5pt]
{\fnfour}&=&\displaystyle{ {h_n\one(\koa)} {\frac{\psi_n\one(\koa)}{\koa}}+{\etar^2}\, {j_n(\koa)}{\frac{\psi_n\three(\koa)}{\koa}}}
\\[5pt]
{\fnfive}&=& \displaystyle{{h_n\one(\koa)} {\frac{\psi_n\one(\koa)}{\koa}}-{j_n(\koa)} {\frac{\psi_n\three(\koa)}{\koa}}}
\\[5pt]
\nonumber
{\fnsix}&=& \displaystyle{{j_n(\koa)} {h_n\one(\koa)} {\frac{\psi_n\one(\kLa)}{\kLa}} {\frac{\psi_n\one(\kRa)}{\kRa}}}
\\[5pt]
\qquad\qquad &&\displaystyle{+{\frac{\psi_n\one(\koa)}{\koa}} {\frac{\psi_n\three(\koa)}{\koa}}{j_n(\kLa)} {j_n(\kRa)}}
\\[5pt]
\nonumber
{\fnseven}&=&
 \displaystyle{\les{h_n\one(\koa)}\ris ^2{\frac{\psi_n\one(\kLa)}{\kLa}} {\frac{\psi_n\one(\kRa)}{\kRa}}}
 \\[5pt]
 \qquad\quad&&
 \displaystyle{+\les{\frac{\psi_n\three(\koa)}{\koa}}\ris^2{j_n(\kLa)}{j_n(\kRa)}}
\end{eqnarray}
and
\begin{eqnarray}
{\fneight} &=&
\displaystyle{-\left(1+{\etar^2}\right)
{h_n\one(\koa)}{\frac{\psi_n\three(\koa)}{\koa}} {\fnone}
+2 {\etar}{\fnseven}}\,.
\end{eqnarray}
\end{subequations}
In these equations, $\etar=\sqrt{\mur}/\sqrt{\epsr}$ is the relative impedance of the
isotropic chiral material, whereas the wavenumbers
\begin{equation}
\left.\begin{array}{l}
\kL=\ko(\sqrt{\epsr}\sqrt{\mur}+\kappa)
\\[5pt]
\kR=\ko(\sqrt{\epsr}\sqrt{\mur}-\kappa)
\end{array}\right\}
\label{def-kLkR}
\end{equation}
depend on the chirality parameter $\kappa$ (in addition to their dependence on $\epsr$ and $\mur$).

Again, in contrast to Secs.~\ref{sec:dms}--\ref{sec:charged}, $\bn$ and $\cn$ are not null valued when the sphere is made
of an isotropic chiral material. Furthermore, since $\bn=\cn$, Eq.~(\ref{Qb-sphere})
informs us that $\Qb$ is definitely independent
of the polarization state of the incident plane wave \cite{Hector}. 

\begin{figure}[!ht]
\centering
\begin{subfigure}{14cm}
    \centering\includegraphics[width=0.8\textwidth]{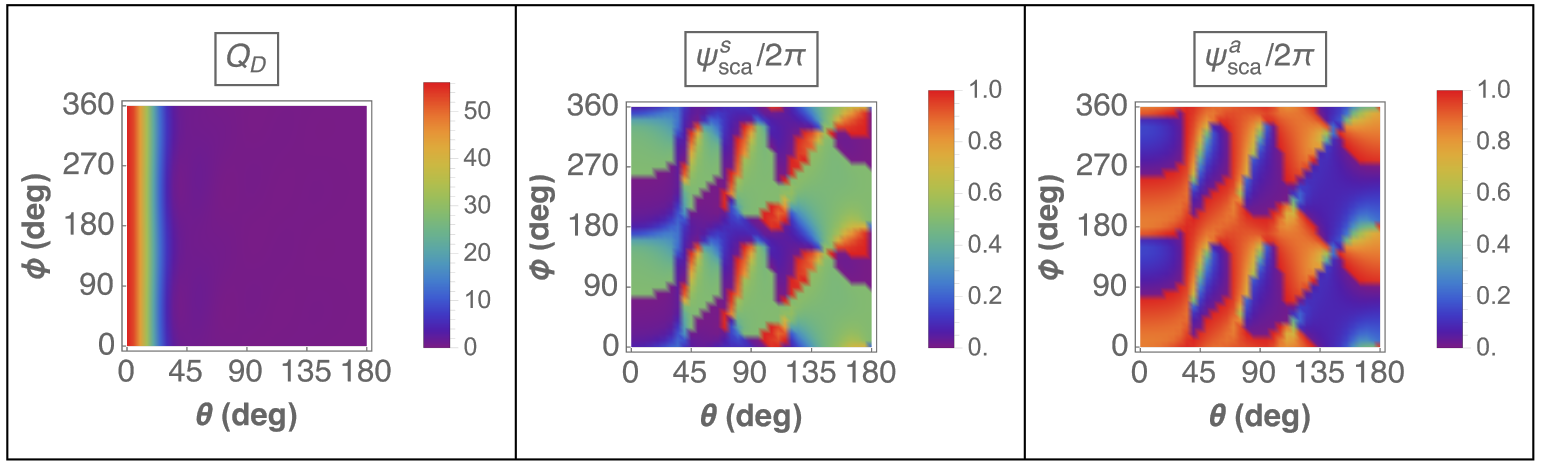}
    \caption{$s$-pol. incidence. $Q\sca=1.8654$, $Q\ext=2.8555$, $\Qf=56.189$, $\Qb=0.0627$.}
    \label{ChiralPos-s}
\end{subfigure}
\hfill
\begin{subfigure}{14cm}
     \centering\includegraphics[width=0.8\textwidth]{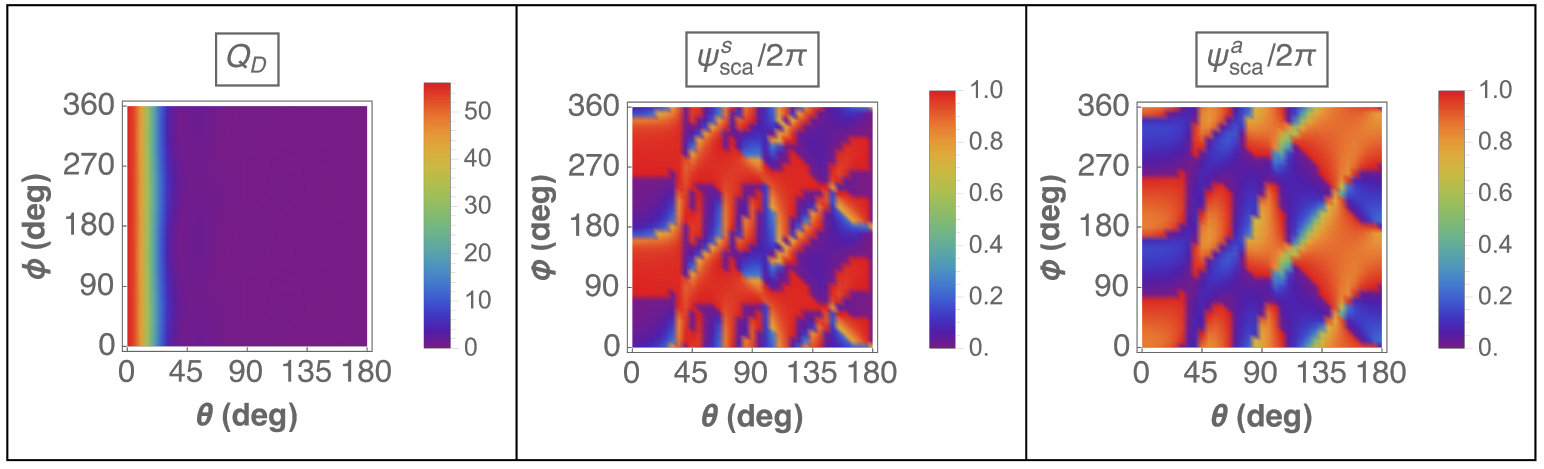}
    \caption{$p$-pol. incidence. $Q\sca=1.8654$, $Q\ext=2.8555$, $\Qf=56.189$, $\Qb=0.0627$.}
    \label{ChiralPos-p}
\end{subfigure}
\hfill
\begin{subfigure}{14cm}
     \centering\includegraphics[width=0.8\textwidth]{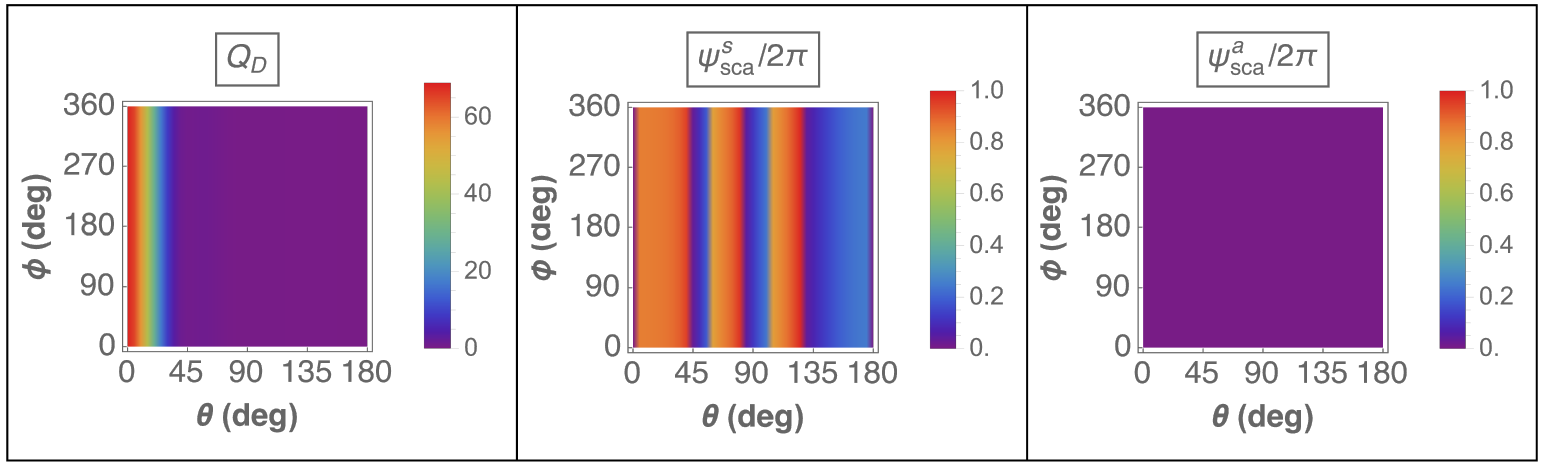}
    \caption{RCP incidence. $Q\sca=1.2080$, $Q\ext=2.6154$, $\Qf=43.529$, $\Qb=0.0627$.}
    \label{ChiralPos-R}
\end{subfigure}
\hfill
\begin{subfigure}{14cm}
     \centering\includegraphics[width=0.8\textwidth]{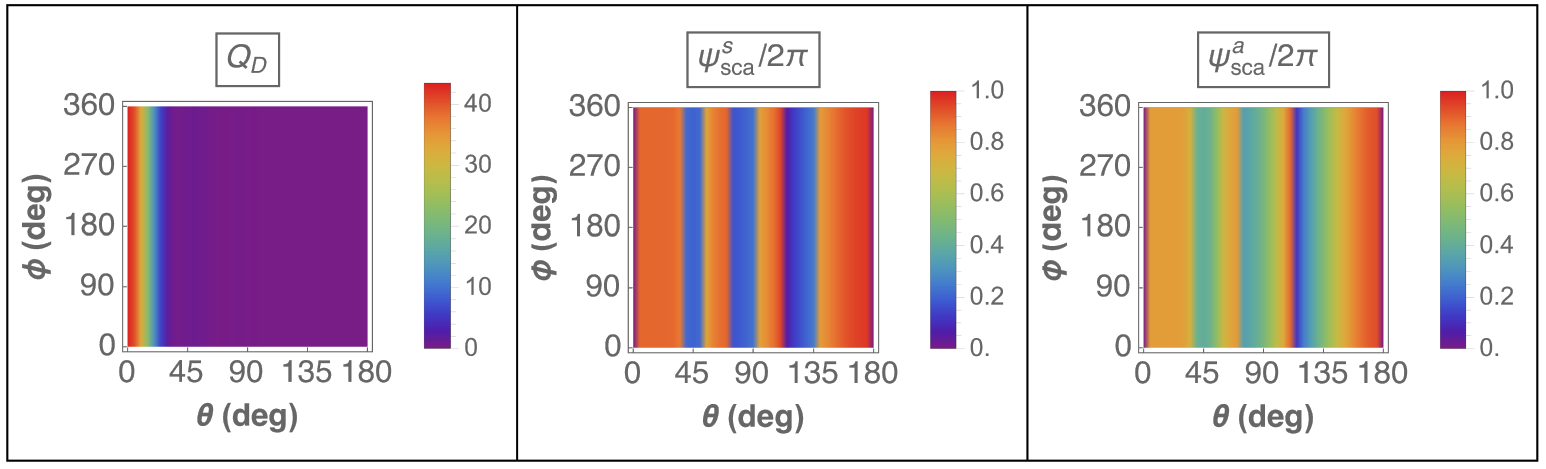}
    \caption{LCP incidence. $Q\sca=2.5228$, $Q\ext=3.0957$, $\Qf=68.850$, $\Qb=0.0627$.}
    \label{ChiralPos-L}
\end{subfigure}
\caption{$\QD$, $\Psis\sca$, and $\Psia\sca$  as functions of $\theta$ and $\phi$ for
an isotropic chiral sphere 
of size parameter $\koa=5$.
The chiral material is characterized by $\epsr=3(1+0.1i)$, $\mur=1.1(1+0.05i)$, and $\kappa=0.5(1+0.2i)$ in Eqs.~(\ref{def-chiral}). }
\label{Fig:ChiralPos} 
\end{figure}

\begin{figure}[!ht]
\centering
\begin{subfigure}{14cm}
    \centering\includegraphics[width=0.8\textwidth]{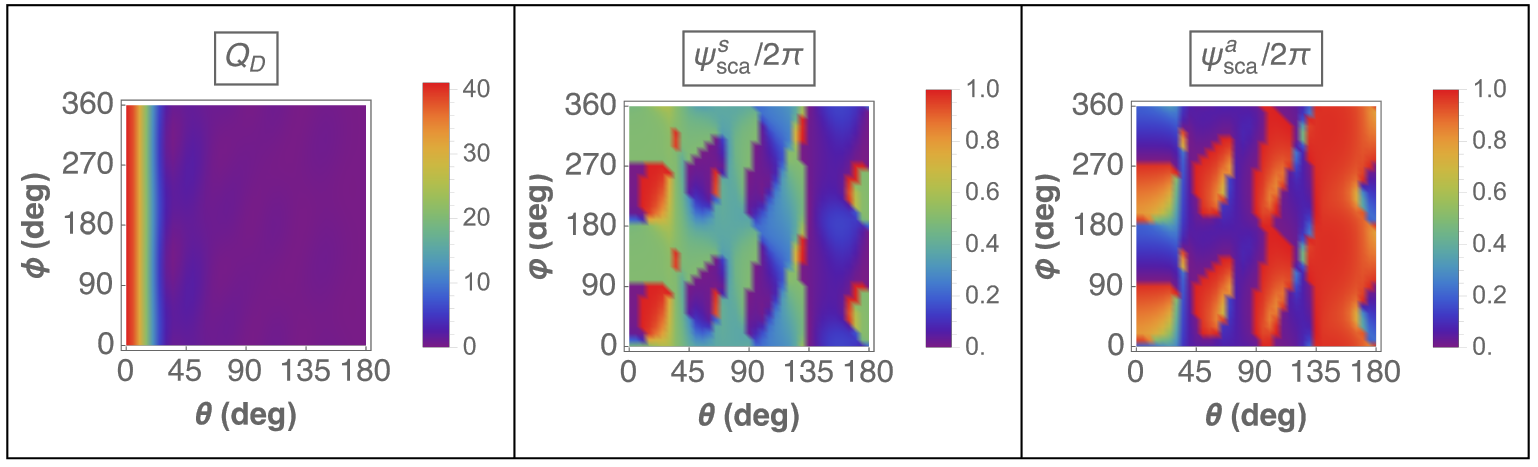}
    \caption{$s$-pol. incidence. $Q\sca=1.4713$, $Q\ext=2.5244$, $\Qf=41.150$, $\Qb=0.0583$.}
    \label{ChiralNeg-s}
\end{subfigure}
\hfill
\begin{subfigure}{14cm}
     \centering\includegraphics[width=0.8\textwidth]{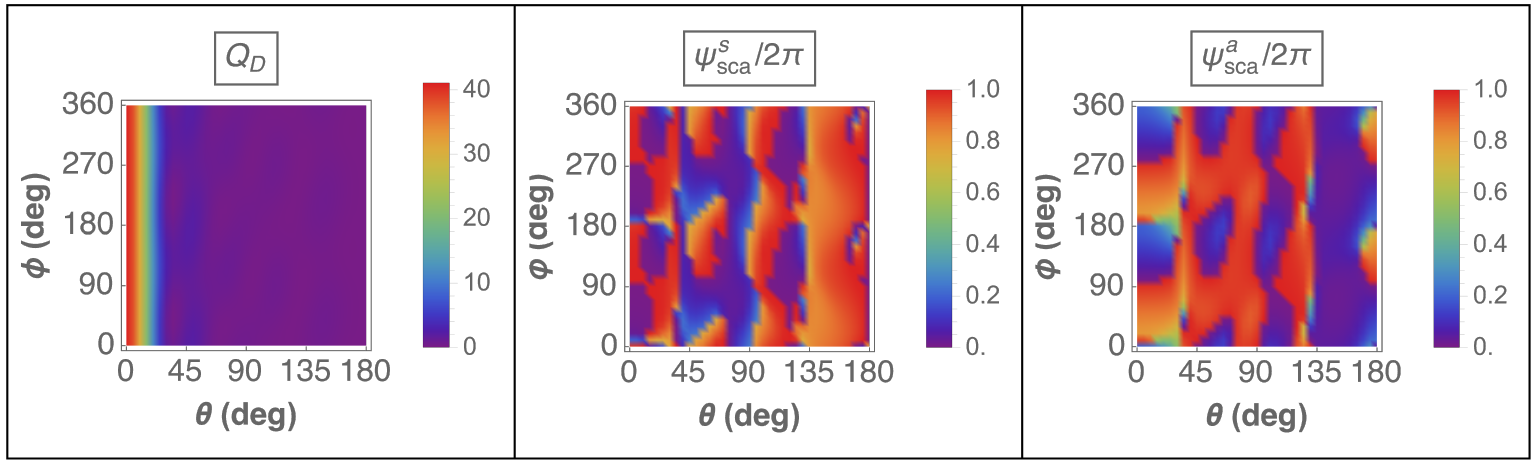}
    \caption{$p$-pol. incidence. $Q\sca=1.4713$, $Q\ext=2.5244$, $\Qf=41.150$, $\Qb=0.0583$.}
    \label{ChiralNeg-p}
\end{subfigure}
\hfill
\begin{subfigure}{14cm}
     \centering\includegraphics[width=0.8\textwidth]{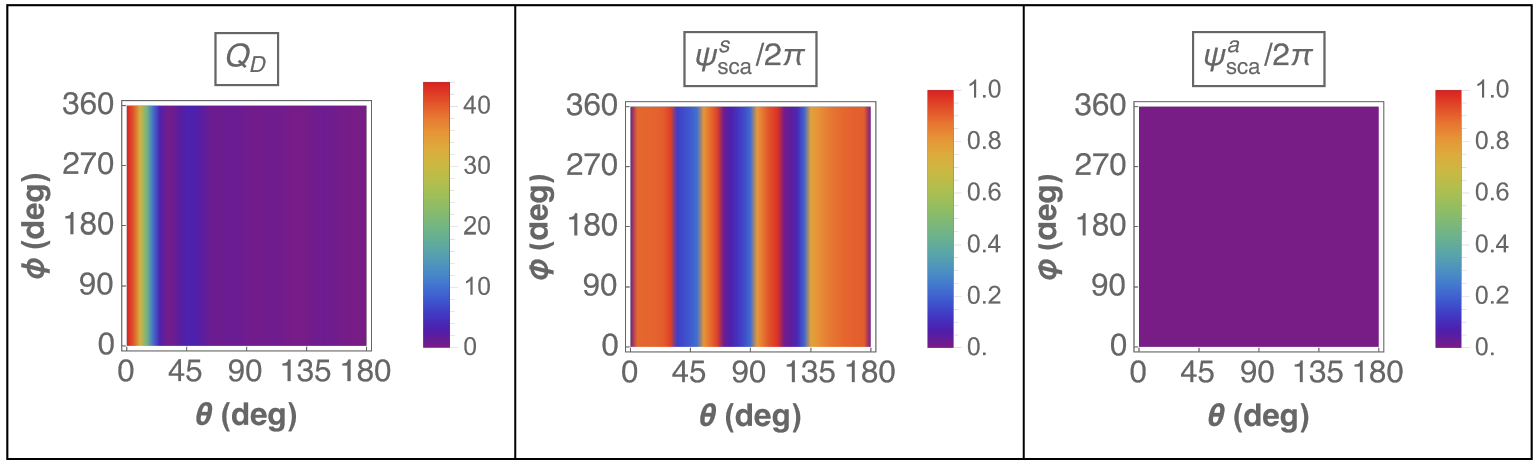}
    \caption{RCP incidence. $Q\sca=1.2308$, $Q\ext=2.4652$, $\Qf=38.329$, $\Qb=0.0583$.}
    \label{ChiralNeg-R}
\end{subfigure}
\hfill
\begin{subfigure}{14cm}
     \centering\includegraphics[width=0.8\textwidth]{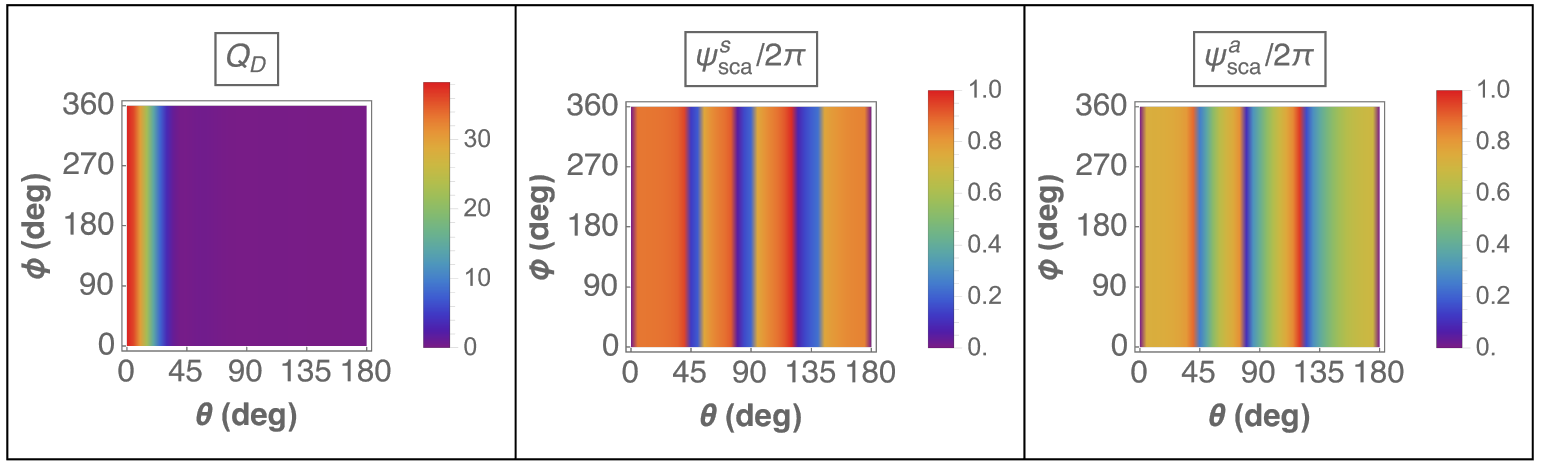}
    \caption{LCP incidence. $Q\sca=1.7117$, $Q\ext=2.5835$, $\Qf=43.970$, $\Qb=0.0583$.
    }
    \label{ChiralNeg-L}
\end{subfigure}
\caption{Same as Fig.~\ref{Fig:ChiralPos} except that $\kappa=0.5(-1+0.2i)$. }
\label{Fig:ChiralNeg} 
\end{figure}

Density plots of $\QD(\ur)$, $\Psis\sca(\ur)$,
and $\Psia\sca(\ur)$ as functions of the scattering angles $\theta$ and
$\phi$ are provided in Fig.~\ref{Fig:ChiralPos} for a sphere of size
parameter $\koa=5$ and made of a chiral material
with $\epsr=3(1+0.1i)$, $\mur=1.1(1+0.05i)$, and $\kappa=0.5(1+0.2i)$.  Figure~\ref{Fig:ChiralNeg} 
provides the same information for $\epsr=3(1+0.1i)$, $\mur=1.1(1+0.05i)$, and $\kappa=0.5(-1+0.2i)$.
 The following remarks emerged   from and examination of these two figures and other data (not shown):
\begin{itemize}

\item \textit{Remark \ref{sec:chi}A}: $\QD(\ur)$, $\Psis\sca(\ur)$, and $\Psia\sca(\ur)$ are dependent on the polarization state
of the incident plane wave, in contrast to $\Qb$. Each of the efficiencies $Q\sca$, $Q\abs$, $Q\ext$, and $\Qf$ is dependent on
that polarization state, with the proviso that it has the same value for incident $s$- and $p$-polarized plane waves.

\item \textit{Remark \ref{sec:chi}B}: $\Psis\sca(\ur)$ depends on both $\theta$ and $\phi$, but only on $\theta$ when the incident plane wave is
circularly polarized.  

\item \textit{Remark \ref{sec:chi}C}: Provided that ${\rm Im}(\kappa)=0$, the values of
$\Psis\sca(\ur)$ for incident RCP and LCP plane waves are interchanged when $\kappa$ is replaced by $-\kappa$.

\item \textit{Remark \ref{sec:chi}D}: $\Psia\sca(\ur)$ depends on both $\theta$ and $\phi$, but only on $\theta$ when the incident plane wave is
LCP.   

\item \textit{Remark \ref{sec:chi}E}: The difference $\vert\Psis\sca(\ur)-\Psia\sca(\ur)\vert$ is $\phi$-independent, when the incident plane wave is
circularly polarized.

\end{itemize}

\subsection{Potential for Use}\label{sec:pu}
A comparison of Figs.~\ref{Fig:DMlossless}--\ref{Fig:ChiralNeg} easily reveals that the density plot of $\QD(\ur)$ has very few features
compared to the density plots of   $\Psis\sca(\ur)$ and  $\Psia\sca(\ur)$. Change the 
constitution---whether isotropic, chiral, anisotropic, or bianisotropic; whether homogeneous or nonhomogeneous---of the 3D object even slightly and the density
plots of $\Psis\sca(\ur)$ and  $\Psia\sca(\ur)$ will change a lot in comparison to the density plot of $\QD(\ur)$. 
Although not illustrated in Secs.~\ref{sec:dms}--\ref{sec:chi}, a change in the size of the scattering object is also marked far more in the density plots of $\Psis\sca(\ur)$ and  $\Psia\sca(\ur)$ than in the density plot of $\QD(\ur)$. Any alteration in the shape of the scatterer from   spherical is expected to be similarly consequential. And then there are the effects of changing the boundary conditions and the polarization state of the incident plane wave on the 3D object. Finally, some linear or even nonlinear combination of the symmetric and the asymmetric geometric spinors may be used to create density plots containing more accessible information about the  scatterer.
The inescapable conclusion at this early stage of research on geometric-phase portrayals of electromagnetic scattering is that
it appears promising for deployment to solve inverse-scattering problems through diverse techniques \cite{Xudong,CK,LLM,class,Maokun}.

\section{Concluding Remark}\label{sec:cr}
 The concept of geometric phase in optics emerged from a need to compare two plane waves \cite{Pancha}, expanded to encompass both classical and quantum-mechanical phenomenons \cite{SWbook,Citro}, and is nowadays  applied to design planar optical devices of certain classes \cite{Brasselet2017,metasurface,Faraz,Jisha}. In this essay, I have initiated the geometric-phase portrayal of electromagnetic scattering by a 3D object embedded in free space. I have shown  that this portrayal is highly sensitive to changes in the shape, size, and composition of that object, boundary conditions, and the polarization state
 of the incident plane wave. As this portrayal is expect to supplement information delivered by the differential scattering efficiency, 
 I hope to have inspired experimental, theoretical, and numerical investigations relevant to forward- and inverse-scattering problems through this essay.

\section*{Appendix 1: Vector Spherical Wavefunctions}
 The vector spherical wavefunctions  \cite{Stratton1941,MF1953}
\begin{subequations}
\begin{equation}
{\#M}\eomn\one(\ko{\#r}) =j_n(\ko r)
\les\utheta \,f\eomn(\theta,\phi)-\uphi \,g\eomn (\theta,\phi)\ris
\label{M1-def}
\end{equation}
and
\begin{eqnarray}
\nonumber
&&
{\#N}\eomn\one(\ko{\#r}) =
\ur \,n(n+1) P_n^m(\cos\theta)\frac{j_n(\ko r)}{\ko r}
{\lec\begin{array}{c}{\cos(m\phi)}\\{\sin(m\phi)}\end{array}\ric}
\\[5pt]
&&\qquad+\frac{\psi_n\one(\ko r)}{\ko r}
\les\utheta\,g\eomn(\theta,\phi)+\uphi\,f\eomn(\theta,\phi)\ris\,
\label{N1-def}
\end{eqnarray}
\end{subequations}
are regular at the origin,
with
\begin{subequations}
\begin{equation}
f\eomn(\theta,\phi)=\mp\pinm(\theta)
{\lec\begin{array}{c}{\sin(m\phi)}\\{\cos(m\phi)}\end{array}\ric}
\label{def-fsmn}
\end{equation}
and
\begin{equation}
g\eomn(\theta,\phi)=\taunm(\theta)
{\lec\begin{array}{c}{\cos(m\phi)}\\{\sin(m\phi)}\end{array}\ric}\,.
\label{def-gsmn}
\end{equation}
\end{subequations}
In these expressions,
\begin{subequations}
\begin{equation}
\psi_n\one(w)= \frac{d}{dw}\left[w\,j_n(w)\right]\,,
\end{equation}
\begin{equation}
\pinm(\theta)=\frac{m P_n^m(\cos\theta)}{\sin\theta}\,,
\end{equation}
\begin{equation}
\taunm(\theta)=\frac{dP_n^m(\cos\theta)}{d\theta}\,,
\end{equation}
\end{subequations}
$j_n(\.)$ denotes the spherical Bessel function of order $n$,
and $P_n^m(\.)$ is the associated
Legendre function of order $n$ and degree $m$.
The vector spherical wavefunctions  \cite{Stratton1941,MF1953}
\begin{subequations}
\begin{equation}
{\#M}\eomn\three(\ko{\#r}) = 
h_n\one(\ko r)\les
\utheta \,f\eomn(\theta,\phi) 
-\uphi \,g\eomn (\theta,\phi)\ris\,
\label{M3-def}
\end{equation}
and
\begin{eqnarray}
\nonumber
&&
{\#N}\eomn\three(\ko{\#r}) =
 \ur \,n(n+1) P_n^m(\cos\theta)\frac{h_n\one(\ko r)}{\ko r}
{\lec\begin{array}{c}{\cos(m\phi)}\\{\sin(m\phi)}\end{array}\ric}
\\[5pt]
&&\qquad+ \frac{\psi_n\three(\ko r)}{\ko r}
\les\utheta\,g\eomn(\theta,\phi)+\uphi\,f\eomn(\theta,\phi)\ris\,
\label{N3-def}
\end{eqnarray}
\end{subequations}
are regular at infinity, with
\begin{equation}
 \psi_n\three(w)= \frac{d}{dw}\left[w\,h_n\one(w)\right]
\end{equation}
and $h_n\one(\.)$ as the spherical Hankel function of the first kind and order $n$. 
Note that ${\#M}_{o0n}^{(j)}(\ko{\#r})\equiv \#0$
and  ${\#N}_{o0n}^{(j)}(\ko{\#r})\equiv \#0$, $j\in\lec1,3\ric$.

\section*{Appendix 2: Two Special Directions}
The direction for which either $\theta=0$ or $\theta=\pi$
is either co-parallel or anti-parallel to the $z$ axis. These two
directions often have special  relevance in scattering problems.
Since
\begin{equation}
\pinm(0)=\taunm(0)=(1/2)n(n+1)\delta_{m1}\,,
\end{equation}
Eqs.~(\ref{def-fsmn}) and (\ref{def-gsmn}) yield
\begin{equation}
\left.\begin{array}{c}
f\eomn(0,\phi)= (1/2) n(n+1)
{\lec\begin{array}{c}{-\sin\phi}\\{\cos\phi}\end{array}\ric}
\delta_{m1}
\\[8pt]
g\eomn(0,\phi)=(1/2) n(n+1)
{\lec\begin{array}{c}{\cos\phi}\\{\sin\phi}\end{array}\ric}
\delta_{m1}
\end{array}\ric\,.
\end{equation}
Similarly, because
\begin{equation}
\pinm(\pi)=-\taunm(\pi)=(-)^{n+1}(1/2)n(n+1)\delta_{m1}\,,
\end{equation}
we have
\begin{equation}
\left.\begin{array}{c}
f\eomn(\pi,\phi)=  (-)^{n} (1/2) n(n+1)
{\lec\begin{array}{c}{\sin\phi}\\{-\cos\phi}\end{array}\ric}
\delta_{m1}
\\[8pt]
g\eomn(\pi,\phi)=(-)^{n}(1/2) n(n+1)
{\lec\begin{array}{c}{\cos\phi}\\{\sin\phi}\end{array}\ric}
\delta_{m1}
\end{array}\ric\,.
\end{equation}

\bigskip
\noindent \textbf{Acknowledgments.} Of the numerous researchers that I have had the good fortune to interact with on electromagnetic scattering  for twoscore years and more,  I must name the following ones (in alphabetical order) for especially insightful discussions: Hamad M. Alkhoori, Craig F. Bohren, Muhammad Faryad, Timothy M. Garner, Magdy F. Iskander, Tom G. Mackay, Sergey A. Maksimenko,  Peter B. Monk,   Gregory Ya. Slepyan, and Miguel A. Solano. And I thank Nikolai G. Khlebtsov, M. Pinar Mengu\c{c}, Gerard Gouesbet, Nikolaos L. Tsitsas, and Ping Yang for encouragement to write this essay.

\medskip\noindent \textbf{Declaration of competing interests.} I have nothing to declare.

\medskip\noindent\textbf{Funding source.} Evan Pugh Professorships Endowment at Penn State.

\medskip\noindent\textbf{Declaration of generative AI use.} None whatsoever.

\medskip\noindent \textbf{Data availability.} I will supply data generated for this paper   to other researchers upon reasonable request.

\end{document}